\documentclass[aps,prd,superscriptaddress,groupedaddress,nofootinbib,nobibnotes]{revtex4}

\usepackage{graphicx}
\usepackage{dcolumn}
\usepackage{bm}
\usepackage{amssymb}
\usepackage{epstopdf}
\usepackage{amsmath}
\usepackage{amsfonts}
\usepackage{hyperref}
\hypersetup{
    urlcolor=cyan,
}
\usepackage{color}
\usepackage{mathrsfs}
\usepackage{float}
\usepackage{wick}
\usepackage{tikz-feynman}
\usepackage{tikz-feynhand}
\usepackage[export]{adjustbox}

\def\smallsum{\mathop{\textstyle\sum}\limits}                 

\newcommand{\be}{\begin{equation}}
\newcommand{\ee}{\end{equation}}
\newcommand{\ba}{\begin{eqnarray}}
\newcommand{\ea}{\end{eqnarray}}
\newcommand{\nn}{\nonumber}
\newcommand{\barr}{\begin{array}}
\newcommand{\earr}{\end{array}}

\newcommand\lsim{\mathrel{\rlap{\lower4pt\hbox{\hskip1pt$\sim$}}
        \raise1pt\hbox{$<$}}}
\newcommand\gsim{\mathrel{\rlap{\lower4pt\hbox{\hskip1pt$\sim$}}
        \raise1pt\hbox{$>$}}}

\def\hv_r{{\hat v_r}}
\def\k{{\bf k}}
\def\q{{\bf q}}
\def\p{{\bf p}}
\def\v{{\bf v}}
\def\x{{\bf x}}
\def\l{{\bf l}}
\def\th{{\boldsymbol\theta}}
\def\L{{\mathcal L}}
\def\tdelta{{\tilde\delta}}
\def\tT{{\tilde T}}
\def\hr{{\bf \hat r}}
\def\hk{{\bf \hat k}}
\def\fnlmed{f_{NL}^{\rm med}}

\begin{document}

\title{Exploring KSZ velocity reconstruction with $N$-body simulations and the halo model}

\author{Utkarsh Giri}
\affiliation{Perimeter Institute for Theoretical Physics, Waterloo, ON N2L 2Y5, Canada}
\affiliation{Department of Physics and Astronomy, University of Waterloo, Waterloo, ON N2L 3G1, Canada}

\author{Kendrick~M.~Smith}
\affiliation{Perimeter Institute for Theoretical Physics, Waterloo, ON N2L 2Y5, Canada}

\date{\today}

\begin{abstract}
KSZ velocity reconstruction is a recently proposed method for mapping 
the largest-scale modes of the universe, by applying a quadratic estimator
$\hv_r$ to the small-scale CMB and a galaxy catalog.
We implement kSZ velocity reconstruction in an $N$-body simulation
pipeline and explore its properties.
We find that the reconstruction noise can be larger than the analytic
prediction which is usually assumed.
We revisit the analytic prediction and find additional noise terms which explain the discrepancy.
The new terms are obtained from a six-point halo model calculation,
and are analogous to the $N^{(1)}$ and $N^{(3/2)}$ biases
in CMB lensing.
We implement an MCMC pipeline which estimates $f_{NL}$ from 
$N$-body kSZ simulations, and show 
that it recovers unbiased estimates of $f_{NL}$, with statistical
errors consistent with a Fisher matrix forecast.
Overall, these results confirm that kSZ velocity reconstruction will be 
a powerful probe of cosmology in the near future, but new terms
should be included in the noise power spectrum.
\end{abstract}

\maketitle

\tableofcontents

\section{Introduction}

The cosmic microwave background (CMB) has been a gold mine
of cosmological information.
So far, the constraining power of the CMB has come mainly
from ``primary'' anisotropy from the last scattering surface
(which dominates at angular wavenumbers $l \lesssim 2000$)
and gravitational lensing (which dominates at 
$2000 \lesssim l \lesssim 4000$).
On even smaller scales ($l \gtrsim 4000$), the CMB temperature
is dominated by the kinetic Sunyaev-Zeldovich (kSZ) effect: Doppler shifting
of CMB photons by free electrons in the late universe.

The kSZ effect has been detected in cross-correlation with 
large-scale structure, with the latest measurements approaching 
10$\sigma$~\cite{Hand:2012ui,Ade:2015lza,Schaan:2015uaa,Soergel:2016mce,Hill:2016dta,DeBernardis:2016pdv,Schaan:2020qhk},
and upcoming experiments such as Simons Observatory~\cite{Ade_2019}
should make percent-level measurements in the next few years.
In anticipation of these upcoming measurements, it is very
interesting to ask how best to constrain cosmological parameters with
the kSZ effect, possibly in cross-correlation with large-scale
structure.

A variety of kSZ-sensitive statistics have been proposed
(e.g.~\cite{Dore:2003ex,DeDeo:2005yr,Ho:2009iw,Hand:2012ui,Li:2014mja,Alonso:2016jpy,Smith:2016lnt,Deutsch:2017ybc,Smith:2018bpn}
and references therein), but in this paper we focus on the velocity reconstruction estimator 
$\hv_r$ from~\cite{Deutsch:2017ybc}.\footnote{The term
``velocity reconstruction'' is sometimes used to refer to
two different statistics. First, the quadratic estimator
$\hv_r$ which reconstructs velocity modes from the kSZ
and large-scale structure. Second, a linear operation which
reconstructs velocity modes from a galaxy catalog (with no
kSZ input, although this operation is an ingredient in a
kSZ stacking analysis~\cite{Schaan:2015uaa,Schaan:2020qhk}).
In this paper, ``velocity reconstruction'' always refers
to the kSZ quadratic estimator $\hv_r$.}
Velocity reconstruction is a particularly convenient kSZ estimator for cosmological applications, since it 
is straightforward to include $\hv_r$ in analyses involving multiple large-scale structure 
fields, or incorporate complications like redshift-space distortions (RSD)~\cite{Smith:2018bpn}.

The kSZ velocity reconstruction $\hv_r$ is a quadratic estimator which reconstructs
the large-scale radial velocity $\hv_r(\k_L)$ from small-scale
modes of a galaxy field $\delta_g(\k_S)$ and CMB temperature $T(\l)$.
(For the precise definition, see~\S\ref{sec:vrec_results}.)
The scales involved are roughly $k_L \sim 10^{-2}$ Mpc$^{-1}$,
$k_S \sim 1$ Mpc$^{-1}$, and $l_{\rm CMB} \sim 5000$.
Thus, the underlying signal for the reconstruction is the velocity
field on large scales where it can be modelled very accurately, but the
reconstruction noise is hard to model, since the noise is derived from
nonlinear scales.

KSZ velocity reconstruction is interesting for cosmology because its
noise power spectrum is smaller on large scales than previously
known methods, such as galaxy surveys, as we will explain in the
next few paragraphs.
First, we note that on large scales, linear theory is a good approximation,
and radial velocity $v_r$, velocity $v$, and matter overdensity
$\delta_m$ are related in Fourier space by:
\be
v_r(\k_L) = \mu v(\k_L) = \mu \frac{faH}{k_L} \delta_m(\k_L)  \label{eq:vr_delta}
\ee
Here, $f = \partial \log D / \partial \log a$ is the usual
RSD parameter, and $\mu = k_{Lr}/k_L$ is the cosine of the angle
between the Fourier mode $\k_L$ and the line of sight.
Therefore, by applying appropriate factors of $\mu$
and $(faH/k)$, the radial velocity reconstruction $\hv_r$
may be viewed as a reconstruction of $v$ or $\delta_m$.
This allows us to compare the noise power spectrum of 
kSZ velocity reconstruction to other LSS observables, which
measure the density field $\delta_m$.

To take a concrete example, consider a galaxy survey, which
measures the density field $\delta_m$ with a noise power spectrum
$N(k) = b_g^{-2} n_g^{-1}$ which is constant on large scales.
The noise power spectrum of the kSZ velocity reconstruction $\hv_r(\k_L)$
is more complicated, but for now we just note that $N_{v_r}(k_L)$ is also 
constant on large scales. (The noise power spectrum $N_{v_r}$ will be discussed
in depth in \S\ref{sec:vrec_results}, \S\ref{sec:noise_contributions}.)
To compare the two, we convert the kSZ velocity reconstruction to a
reconstruction of $\delta_m$ using Eq.~(\ref{eq:vr_delta}), obtaining
noise power spectrum:
\be
N_{\delta_m}^{\rm kSZ}(\k_L) = \left( \frac{k_L}{\mu f a H} \right)^2 N_{v_r}(\k_L)  \label{eq:N_delta}
\ee
Due to the factor $k_L^2$ on the RHS, the kSZ-derived reconstruction
of the large-scale modes has {\em parametrically lower noise} than the
galaxy field.\footnote{Loophole: This is only true for modes where $|\mu|=|k_r|/k$
is not too small. For modes with small $\mu$, the factor $\mu^{-2}$
in Eq.~(\ref{eq:N_delta}) acts as an SNR penalty, and ``transverse''
modes with $\mu=0$ cannot be reconstructed at all from the kSZ.} 
This low-noise large-reconstruction has several potential
applications (e.g.~\cite{Mueller_2015,Zhang:2015uta,Terrana_2017,Cayuso:2019hen,Hotinli:2019wdp,Contreras:2019bxy}),
but we will concentrate on the cosmological parameter $f_{NL}$.
In~\cite{Munchmeyer:2018eey,Contreras:2019bxy}, it was shown that
adding kSZ data to an analysis of galaxy clustering can significantly
improve $f_{NL}$ constraints, relative to the galaxies alone.
In this forecast, the $f_{NL}$ sensitivity arises from non-Gaussian
bias~\cite{Dalal:2007cu,Slosar:2008hx} in the galaxy survey.
The field $v_r$ is not directly sensitive to $f_{NL}$, but including it 
helps improve the $f_{NL}$ constraint, using the idea
of sample variance cancellation~\cite{Seljak_2009}.

Summarizing, kSZ velocity reconstruction estimator
is emerging as an interesting new tool for constraining cosmology,
using upcoming kSZ and large-scale structure data.
However, there is currently a major caveat.
As mentioned above, the reconstruction $\hv_r$ is derived from LSS modes 
on scales $k_S \sim 1$ Mpc$^{-1}$, and therefore the reconstruction noise
depends on statistics of nonlinear modes which
are difficult to model. Forecasting work so far (e.g.~\cite{Deutsch:2017ybc,Smith:2018bpn,Munchmeyer:2018eey})
has used simple analytic models which approximate the true statistics of
the reconstruction noise.
The purpose of this paper is to assess the validity of these approximations, by applying 
kSZ velocity reconstruction to $N$-body simulations.

In the bullet points below, we separate the issues by dissecting the different
approximations which are usually made, and summarize the main results of this paper.
\begin{itemize}
\item In~\cite{Smith:2018bpn} it was argued that the kSZ-derived 
 velocity reconstruction $\hv_r(\k)$ is an estimator of the true radial
 velocity on large scales:
 \be
 \hv_r(\k) = b_v v_r^{\rm true}(\k) + \mbox{(reconstruction noise)}
 \ee
 where the reconstruction noise is uncorrelated with $v_r^{\rm true}$,
 and the bias $b_v$ is constant on large scales.
 The value of $b_v$ depends on the mismatch between the true small-scale
 galaxy-electron power spectrum $P_{ge}(k_S)$ and the fiducial spectrum
 $P_{ge}^{\rm fid}(k_S)$ used to construct the quadratic estimator.
 
 In this paper, we will confirm all of these ``map-level'' properties
 of the velocity reconstruction estimator $\hv_r$ using $N$-body simulations.
 We also find that, although $b_v$ is constant on the largest scales,
 it starts to acquire scale dependence at a surprisingly small value of $k$
 (see Figure~\ref{fig:quijote_correlation_and_bias}).

\item Moving from ``map level'' to power spectra, we next consider the
 power spectrum of the reconstruction noise.
 In~\cite{Smith:2018bpn}, an analytic model was given for the 
 noise power spectrum, which makes the approximation that the small-scale
 galaxy field $\delta_g(\k_S)$ and the small-scale CMB $T(\l)$ are
 uncorrelated.
 This is a good approximation if the CMB modes are noise-dominated,
 but potentially dubious if the CMB is kSZ-dominated.
 In this paper, we will denote the reconstruction noise power spectrum 
 computed in this approximation by $N_{v_r}^{(0)}(\k_L)$, and call it
 the ``kSZ $N^{(0)}$-bias''.
 This terminology is intended to emphasize an analogy with CMB lensing
 which will be explained later in the paper.
 
 We compare the reconstruction noise power spectrum in our simulations
 with the kSZ $N^{(0)}$-bias, and find a significant discrepancy,
 even in the limit $\k_L \rightarrow 0$.
 For the fiducial survey parameters used in this paper (see~\S\ref{sec:pipeline}), 
 the $N^{(0)}$-bias underpredicts the true reconstruction noise
 power spectrum by a factor 2--3.
 This turns out to have a small effect on the bottom-line constraint
 on $f_{NL}$, but this may not be the case for other choices of survey parameters
 (CMB noise, galaxy density, redshift, etc.)
 This result shows that the $N^{(0)}$-bias proposed in~\cite{Smith:2018bpn} as
 a model for reconstruction noise is sometimes incomplete.

\item
 Motivated by this discrepancy between theory and simulation,
 we revisit the calculation of the kSZ reconstruction noise,
 and find additional terms. The new terms are analogous to
 the $N^{(1)}$-bias~\cite{Kesden:2003cc} and 
 $N^{(3/2)}$-bias~\cite{Bohm:2016gzt} in CMB lensing,
 and are obtained from a six-point halo model calculation.
 We calculate the new terms under some simplifying approximations,
 and find that they explain the excess noise seen
 in simulations (see Figure~\ref{fig:N3/2 bias}).
 The new terms are algebraically simple enough
 that including them in future forecasts or data analysis should
 be straightforward (see Eq.~(\ref{eq:n32_final})).
 
\item Moving from the reconstruction noise power spectrum to
 higher-point statistics, we next study the question of whether
 reconstruction noise is a Gaussian field.
 As a simple test for Gaussianity, we compute the correlation
 matrix between $k$-bands of the estimated reconstruction noise
 power spectrum (which would be the identity matrix
 for a Gaussian field).
 The bandpower covariance determines
 statistical errors on parameters derived from power spectra.
 In particular, the $f_{NL}$ Fisher matrix forecasts from~\cite{Munchmeyer:2018eey}
 implicitly assume that bandpower correlations are small, and we
 would like to test this assumption.
 
 In simulation,
 we find that bandpower correlations are small on the very large scales 
 which dominate $f_{NL}$ constraints, but increase rapidly with $k$, 
 and become order-one at $k \sim 0.03$ Mpc$^{-1}$.
 
\item Putting everything together, we develop an ``end-to-end''
 pipeline which recovers $f_{NL}$ from a simulated galaxy catalog
 and kSZ map. The pipeline applies the quadratic estimator $\hv_r$,
 then performs MCMC exploration of the posterior likelihood for
 parameters $(f_{NL}, b_g, b_v)$, given realizations of the
 galaxy field $\delta_g$ and velocity reconstruction $\hv_r$.
 When deriving the posterior likelihood, we assume that 
 the reconstruction noise power spectrum is equal to the
 $N^{(0)}$-bias, and that the reconstruction noise is a Gaussian field.
 Based on previous bullet points, these approximations are
 imperfect, but their impact on parameter constraints should be
 small, and therefore it seems plausible that the posterior likelihood
 will produce valid parameter constraints.
 
 We find that kSZ velocity reconstruction works!
 We run the pipeline on simulations with both zero and
 nonzero $f_{NL}$, in a noise regime where sample variance cancellation
 is important, and demonstrate that it recovers unbiased $f_{NL}$
 estimates, with statistical errors consistent with Fisher matrix
 forecasts.
\end{itemize}

These results largely serve as zeroth-order validation of
the basic kSZ velocity reconstruction framework
from~\cite{Deutsch:2017ybc,Smith:2018bpn} and $f_{NL}$
forecasts from~\cite{Munchmeyer:2018eey}, with the
addition of new terms in the reconstruction noise.
This initial exploratory study can be extended in several
interesting directions; see~\S\ref{sec:discussion} for
systematic discussion.

To test kSZ velocity reconstruction as accurately as possible, we want to
use as much simulation volume as we can.
For this reason, we use collisionless $N$-body simulations, which have much
lower computational cost per unit volume than hydrodynamical simulations.
We approximate the electron overdensity field by the dark matter field
($\delta_e = \delta_m$), and approximate the galaxy catalog by a halo
catalog ($\delta_g = \delta_h$).
These are crude approximations, and in particular our approximation
$\delta_e = \delta_m$ means that we overpredict the galaxy-electron
power spectrum $P_{ge}(k_S)$ by an order-one factor.
However, in this paper our goal is to compare theory and simulation, and
the level of agreement is unlikely to depend on details of small-scale 
power spectra, as long as the analysis is self-consistent.
Since we use collisionless simulations, we can also leverage the
high-resolution Quijote public simulations~\cite{Villaescusa-Navarro:2019bje} 
with a total volume of 100 Gpc$^3$ and $f_{NL}=0$.
For $f_{NL} \ne 0$, we run \texttt{GADGET-2}~\cite{Springel:2005mi} with a
custom initial condition generator.

This paper builds on previous papers which explore the effects of
primordial non-Gaussianity in $N$-body simulations,
e.g.~\cite{Dalal:2007cu,Desjacques_2009,Pillepich:2008ka,Grossi:2009an,Giannantonio:2009ak,Wagner:2011wx,Hamaus_2011,Scoccimarro:2011pz,Baldauf:2015vio,Biagetti:2016ywx} 
and references therein.
The new ingredient is the kSZ velocity reconstruction $\hv_r$.
To our knowledge, there is only one previous paper which explores
kSZ velocity reconstruction in simulations~\cite{Cayuso_2018}.
There, a large correlation was found between 
the reconstructed radial velocity $\hv_r$ and the true radial velocity
$v_r^{\rm true}$, but the reconstruction noise was not compared with
theory, and non-Gaussian simulations were not studied.

This paper is organized as follows.
In~\S\ref{sec:pipeline}, we describe our simulation pipeline
for generating large-scale structure and kSZ realizations.
In~\S\ref{sec:power_spectrum_results}, we show large-scale
structure and CMB power spectra from our simulations.
In~\S\ref{sec:vrec_results}, we study the kSZ velocity reconstruction
estimator $\hv_r$ in detail, and characterize key properties such
as bias, noise, and non-Gaussian bandpower covariance.
In~\S\ref{sec:noise_contributions}, we calculate kSZ reconstruction
noise in the halo model, and
find new terms $N^{(1)}$ and $N^{(3/2)}$ which agree with the
simulations.
We present our MCMC-based $f_{NL}$ pipeline in~\S\ref{sec:fnl_results},
and conclude in~\S\ref{sec:discussion}.
The code for this work can be accessed at \url{https://github.com/utkarshgiri/kineticsz}.

\section{Preliminaries and notation}
\label{sec:preliminaries}

\subsection{``Snapshot'' geometry}
\label{ssec:pipeline_geometry}

Following~\cite{Smith:2018bpn}, we use the following simplified ``snapshot'' geometry
throughout the paper. We take the universe to be a periodic 3-d box with comoving side
length $L=1$ $h^{-1}$ Gpc and volume $V=L^3$, ``snapshotted'' at redshift $z_*=2$,
corresponding to comoving distance $\chi_* \approx 5200$ Mpc.
The notation $(\cdot)_*$ means ``evaluated at redshift $z_*$'',
e.g.~$H_*$ is the Hubble expansion rate at $z_*$,
and $\chi_*$ is comoving distance between $z=0$ and $z=z_*$.

Three-dimensional large-scale structure fields, such as the galaxy overdensity $\delta_g(\x)$,
are defined on a 3-d periodic box of comoving side length $L$.
Two-dimensional angular fields, such as the CMB $T(\th)$, are defined on a 2-d periodic
flat sky with angular side length $L/\chi_*$.
We define line-of-sight integration by projecting the 3-d box onto the xy-face of the cube,
with a factor $1/\chi_*$ to convert from spatial to angular coordinates.
We denote transverse coordinates of the box by $(x,y)$, but denote the radial coordinate
by $r$ (not $z$, to avoid notational confusion with redshift).
We denote a unit three-vector in the radial direction by $\hr$,
and denote the transverse part of a three-vector $\x$ by $\x_\perp$.
Thus a galaxy at spatial location $\x$ appears at angular sky location 
$\th=\x_\perp/\chi_*$.

In the full lightcone geometry, the kSZ temperature anisotropy is given by
a line-of-sight integral $T(\th) = \int dr \, K(r) \, (\hr \cdot \q_e(\th,r))$, 
where $\q_e = (1+\delta_e) \v_e$ is the dimensionless electron momentum field, 
and $K(\cdot)$ is the kSZ radial weight function:
\be
K(z) = -T_{\rm CMB} \, \sigma_T \, n_{e,0} \, x_e(z) \, e^{-\tau(z)} \, (1+z)^2
\ee
In the snapshot geometry, this line-of-sight integral becomes:
\be
T_{\rm kSZ}(\th) = K_* \int_0^L dr \, \Big( \hr \cdot \q_e(\chi_*\th + r\hr) \Big)  \label{eq:ksz_los}
\ee

\subsection{Fourier conventions}

Our Fourier conventions for a 3-d field $f(\x)$ with power spectrum $P(k)$ are:
\be
f(\k) = \int d^3\x \, f(\x) e^{-i\k\cdot\x}
  \hspace{1cm}
\langle f(\k) f(\k')^* \rangle = P(k) (2\pi)^3 \delta^3(\k-\k')  \label{eq:fourier_3d}
\ee
In a finite pixelized 3-d volume $V$, we use Fourier conventions:
\be
f(\k) = \frac{V}{N_{\rm pix}} \sum_{\x} f(\x) e^{-i\k\cdot\x}
  \hspace{1cm}
\big\langle f(\k) f(\k')^* \big\rangle = V P(k) \delta_{\k\k'}
\label{eq:fourier_3d_finite_vol}
\ee
With these conventions, the radial velocity $v_r(\k)$ and matter overdensity $\delta_m(\k)$
are related in linear theory by:
\be
v_r(\k) = i k_r \left( \frac{faH}{k^2} \right) \delta_m(\k)  \label{eq:vr_lin}
\ee
Here, $f(z) = (\partial\log D(z) / \partial\log a)$, where $D(z)$ is the growth function.

Similarly, our Fourier conventions for a 2-d flat-sky field $f(\th)$ with
angular power spectrum $C_l$ are:
\be
f(\l) = \int d^2\th \, f(\th) e^{-i\l\cdot\th}
  \hspace{1cm}
\big\langle f(\l) f(\l')^* \big\rangle = C_l (2\pi)^2 \delta^2(\l-\l')
\ee
In finite pixelized 2-d area $A$ this becomes:
\be
f(\l) = \frac{A}{N_{\rm pix}} \sum_{\th} f(\th) e^{-i\l\cdot\th}
  \hspace{1cm}
\big\langle f(\l) f(\l')^* \big\rangle = 
A C_l \delta_{\l\l'}  \label{eq:fourier_2d_finite_vol}
\ee
In our code, we often represent 2-d fields using dimensionful coordinates
$\x_\perp = \chi_* \th$ and $\k_\perp = \l / \chi_*$, which eliminates
factors of $\chi_*$ in some equations. For example, the line of sight
integral~(\ref{eq:ksz_los}) becomes 
$T(\x_\perp) = K_* \int dr \, (\hr \cdot q_e(\x_\perp + r\hr)$.

\subsection{Primordial non-Gaussianity and halo bias}

Single-field slow-roll inflation is arguably the simplest model of the early universe.
In this model, the initial curvature perturbation $\zeta$ is a Gaussian field to an excellent approximation~\cite{Acquaviva:2002ud,Maldacena:2002vr}.
This is not the case in many alternative models, and searching for primordial non-Gaussianity
(deviations from Gaussian initial conditions) is a powerful probe of physics of the
early universe.
A wide variety of observationally distinguishable non-Gaussian models has been proposed
(see e.g.~\cite{Akrami:2019izv} and references therein).

In this paper, we will concentrate on ``local-type'' non-Gaussianity, in which the
initial curvature perturbation $\zeta$ is of the form:
\be
\zeta(\x) = \zeta_G(\x) + \frac{3}{5} f_{NL} \big( \zeta_G(\x)^2 - \langle \zeta_G^2 \rangle \big)
\ee
where $\zeta_G$ is a Gaussian field, and $f_{NL}$ is a cosmological parameter to 
be constrained from observations.
Local-type non-Gaussianity is fairly generic in multifield early universe models,
such as curvaton models~\cite{Linde:1996gt,Lyth:2001nq,Lyth:2002my},
or modulated reheating models~\cite{Dvali:2003em,Kofman:2003nx}.
Conversely, there are theorems~\cite{Maldacena:2002vr,Creminelli:2004yq}
which show that $f_{NL}=0$ in single-field early universe models,
i.e.~models in which a single field both dominates the stress-energy
of the early universe, and determines the initial curvature perturbation.

In a pioneering paper~\cite{Dalal:2007cu}, Dalal et al showed that large-scale clustering
of dark matter halos depends sensitively on $f_{NL}$.
More precisely, the halo bias $b_h$ is scale dependent on large scales, with functional form:
\be
b_h(k) = b_g + f_{NL} \frac{b_{ng}}{\alpha(k,z)}  \label{eq:halo_bias_ng}
\ee
where $b_g$ is the Gaussian (scale-independent) bias, and:
\be
\alpha(k,z) \equiv \frac{2 k^2 T(k) D(z)}{3 \Omega_m H_0^2}
\ee
The quantity $\alpha(k,z)$ relates the matter overdensity $\delta_m(\k,z)$ to initial
curvature $\zeta(\k)$ in linear theory: $\delta_m(\k,z) = (3/5) \alpha(k,z) \zeta(\k)$.
On large scales $k\rightarrow 0$, $\alpha(k,z)$ is proportional to $k^2$, leading to
an $f_{NL} k^{-2}$ term in the halo bias.
Thanks to this term, even small values of $f_{NL}$ can produce large observable effects
on large scales.
Although current large-scale structure constraints on $f_{NL}$~\cite{Castorina:2019wmr}
are not competitive with CMB constraints~\cite{Akrami:2019izv},
future LSS experiments which probe large volumes and high redshifts
should be comparable or better than the 
CMB~\cite{ Carbone:2010sb,dePutter:2014lna,Dore:2014cca,Alonso:2015sfa,Ferraro:2014jba,Schmittfull:2017ffw}.

The parameter $b_{ng}$ in Eq.~(\ref{eq:halo_bias_ng}) is given exactly 
by~\cite{Slosar:2008hx,Baldauf:2011bh}:
\be
b_{ng} = 2 \frac{\partial\log n_h}{\partial \log \sigma_8}  \label{eq:bng_exact}
\ee
This exact expression is of limited usefulness, since the derivative on the RHS
is not an observable quantity.
Treating $b_{ng}$ as a free parameter is not a viable option for 
data analysis, since it would be degenerate with $f_{NL}$ (only the combination
$b_{ng} f_{NL}$ would be observable).
However, in  spherical collapse models of halo formation,
$b_{ng}$ is related to the Gaussian bias as:
\be
b_{ng} = 2 \delta_c (b_g - 1)  \label{eq:bng_deltac}
\ee
where $\delta_c$ is the collapse threshold, given by $\delta_c = 1.69$ in the Press-Schechter
model~\cite{Press:1973iz}, or $\delta_c = 1.42$ in Sheth-Tormen~\cite{Sheth:2001dp}.
Although Eq.~(\ref{eq:bng_deltac}) is an approximation to the exact result~(\ref{eq:bng_exact}),
it is usually accurate at the 10--20\%
level~\cite{Desjacques_2009,Pillepich:2008ka,Grossi:2009an,Giannantonio:2009ak,Wagner:2011wx,Hamaus_2011,Scoccimarro:2011pz,Baldauf:2015vio,Biagetti:2016ywx}, and is suitable for data analysis, since the parameters $(b_g, f_{NL})$
can be jointly constrained without degeneracy.

In our $N$-body simulations, we find that Eq.~(\ref{eq:bng_deltac}) gives a good fit to 
the non-Gaussian bias observed in our $N$-body simulations, if the Sheth-Tormen 
threshold $\delta_c = 1.42$ is used. (See Figure~\ref{fig:shot_noise_and_bias} below.)
This is consistent with previous simulation-based
studies~\cite{Pillepich:2008ka,Grossi:2009an,Hamaus_2011,Baldauf:2015vio}, which used the parameterization
$b_{ng} = 2 \sqrt{q} (1.69) (b_g-1)$, and found a fudge-factor $\sqrt{q}$ around 0.84 for
friends-of-friends halos (which we use in our pipeline, see~\S\ref{ssec:nbody_sims}).
In the rest of the paper, we model large-scale halo bias using Eq.~(\ref{eq:halo_bias_ng}),
where $b_{ng}$ is given by Eq.~(\ref{eq:bng_deltac}) with $\delta_c = 1.42$.

\section{Simulation pipeline}
\label{sec:pipeline}

\subsection{Collisionless approximation}
\label{ssec:pipeline_approximation}

Simulating high-fidelity kSZ maps for velocity reconstruction is very 
computationally challenging.
KSZ anisotropy appears on small angular scales in the CMB, where it is sourced by
electron density fluctuations on small scales $k_S \sim 1$ Mpc$^{-1}$,
leading to high resolution requirements in a simulation.
Furthermore, on these small scales, collisionless $N$-body simulations are
not really accurate enough to simulate the electron density, and hydrodynamical
simulations should be used instead, which are much more expensive.
At the same time, the cosmological constraining power of the kSZ comes from
the largest scales, so a large simulation volume is required, if the goal
is to make a simulation with an interesting $f_{NL}$ constraint.
This combination of volume and resolution requirements presents
a serious computational challenge, and new simulation methods are probably
required to satisfy all requirements strictly.

In this paper, our goal is simply to test kSZ velocity reconstruction for biases as 
precisely as possible, under a self-consistent set of assumptions.
For this purpose, perfectly accurate kSZ simulations are not required, and approximations
are acceptable, as long as they are self-consistent.
We will make the approximation that the electron density perfectly traces the dark
matter density ($\delta_e=\delta_m$).
This overestimates power spectra such as $P_{ge}$, $P_{ee}$, or $C_l^{\rm kSZ}$
by an order-one factor on small scales, since hydrodynamic effects suppress
electron fluctuations relative to dark matter~\cite{shaw}.
However, the question of whether kSZ velocity reconstruction is biased is
unlikely to depend on the details of these small scale power spectra.
For our purposes, what is crucial is that the approximation $\delta_e \approx \delta_m$
is applied consistently throughout the simulation and reconstruction pipelines.

The approximation $\delta_e \approx \delta_m$ dramatically
decreases computational cost, since we can use collisionless $N$-body
simulations.
Similarly, instead of simulating galaxies, we use dark
matter halos as a proxy for galaxies, i.e.~we make the
approximation $\delta_g \approx \delta_h$.
In the rest of the paper, we use ``galaxies'' synonymously
with ``halos'', and ``electrons'' synonymously with ``dark matter particles''.

\subsection{$N$-body simulations}
\label{ssec:nbody_sims}

We are interested in collisionless $N$-body simulations for both
zero and nonzero $f_{NL}$.
For $f_{NL}=0$, rather than running our own simulations from scratch,
we use the Quijote simulations~\cite{Villaescusa-Navarro:2019bje},
a large suite of publicly available $N$-body simulations.
We use 100 simulations with $1024^3$ particles and volume 1 $h^{-3}$ Gpc$^3$ each.

For $f_{NL} \ne 0$, we generated a limited number of $N$-body simulations
by running \texttt{GADGET-2}~\cite{Springel:2005mi} with non-Gaussian initial conditions
as follows.
We simulate the initial curvature $\zeta$, by simulating a Gaussian field
$\zeta_G$, and then adding a quadratic term:
\begin{align}
\zeta(\x) = \zeta_G(\x) + \frac{3}{5} f_{NL} \big( \zeta_G(\x)^2 - \langle \zeta_G^2 \rangle \big)
\end{align}
where the squaring operation is performed in real space. 
We evolve $\zeta$ to the Newtonian potential $\Phi$ at redshift $z_{\rm ini}=127$,
using linear transfer functions computed using \texttt{CLASS}~\cite{Lesgourgues:2011re}.
We then generate initial conditions for \texttt{GADGET-2} at $z_{\rm ini}=127$
using the Zeldovich approximation \cite{jeongPhD}:
\begin{align}
    \Psi_i(\q) &= - \partial_i \partial^{-2} \delta_m(\q) 
        = - \frac{2}{3 a^2 H(a)^2} \Big( \partial_i \Phi(\q) \Big) \\
    v_i(\q) &= \frac{\partial \Psi_i}{\partial\tau}
       = -\frac{2}{3aH(a)} \Big( \partial_i \Phi(\q) \Big)
\end{align}
Here, $\q$ is the initial Lagrangian location of particles which in our case occupy
center of 3D mesh, $\Psi_i(\q)$ is the initial particle displacement, and $v_i(\q)$
is the initial velocity.
We evolve particles from $z_{\rm ini}=127$ to $z_*=2$ using \texttt{GADGET-2} with the
same parameters (cosmological parameters, force softening length, etc.) as the
Quijote simulations.

\subsection{Large-scale structure fields: $\delta_m$, $\delta_h$, $q_r$}
\label{ssec:lss_fields}

The output of an $N$-body simulation is a catalog of particles with velocities.
In this section, we describe our postprocessing
of the catalog, to obtain pixelized 3-d maps of
the matter overdensity
$\delta_m(\x)$, halo overdensity $\delta_h(\x)$, and radial momentum $q_r(\x)$.

To compute $\delta_m(\x)$, we grid particle positions on a regular 3D mesh
using the cloud-in-cell (CIC) algorithm~\cite{1981csup.book.....H},
implemented in the public code \texttt{nbodykit}~\cite{Hand:2017pqn}.
We use a 3D mesh with $1024^3$ pixels, corresponding to pixel size 1 $h^{-1}$ Mpc. 

To obtain a halo catalog, we run the \texttt{Rockstar} halo-finder~\cite{Behroozi:2011ju}
on the particle positions. (Note that the Quijote simulations include a halo catalog,
but we run our own halo finder instead, so that simulations with zero and nonzero $f_{NL}$
are processed consistently.)
Rockstar implements a modified version of the Friends-of-Friends (FOF) algorithm~\cite{Davis:1985rj}.
After the halo catalog is produced, it is processed to obtain a halo overdensity map $\delta_h(\x)$
by CIC-gridding halo positions.

We use an FOF linking length of 0.28 and require a minimum of 40 particles
to classify a structure as a halo.
This results in halo bias $b_h \sim 3.24$ and density $n_h^{3d} \approx 2.5\times10^{-4} \  \mathrm{Mpc^{-3}}$ 
for simulations with $f_{nl}=0$ at redshift $z_{*}=2$. 
Since we are using halos as proxies for galaxies, our effective 2-d
galaxy number density is $dn_g^{2d}/dz = (\chi_*^2/H_*) n_h^{3d} = 0.8$
arcmin$^{-2}$. In comparison, DESI has a combined (ELG+LRG+QSO) number density $dn_g^{2d}/dz = 0.91$ 
arcmin$^{-2}$ at its peak at $z=0.75$ \cite{desi2016},
while Vera Rubin Observatory ``gold'' sample will have a number density
$dn_g^{2d}/dz = 36$ arcmin$^{-2}$ at its peak at $z=0.6$ \cite{lsst2009}. 

The radial momentum field deserves some discussion.
We are interested in making 3-d maps of the true radial velocity $v_r^{\rm true}$,
in order to compare it to the kSZ velocity reconstruction $\hat v_r$ on large scales.
However, in an $N$-body simulation, the definition of $v_r^{\rm true}$
is ambiguous. Here are three possibilities:
\begin{enumerate}
  \item We can use the radial momentum $q_r = (1+\delta) v_r$. Since momentum is
    particle-weighted, it can be directly computed from particle positions and
    velocities.
  \item We can use the linear velocity field $v_r^{\rm lin}(\k)$, obtained by applying
    linear transfer functions to the initial conditions.
  \item We can choose a smoothing scale, and define the velocity to be the
    smoothed momentum, divided by the smoothed density (appropriately regulated
    to avoid dividing by zero in voids).
\end{enumerate}
We actually tried all three possibilities, and found that the first (the radial
momentum $q_r$) has the highest correlation with the kSZ velocity reconstruction 
$\hat v_r$.
This makes sense intuitively by considering the case of a ``near-void'' region
whose density is close to zero.
In a near-void region, the velocity reconstruction $\hat v_r$ is small, since a factor
of the small-scale inhomogeneity $\delta_g(\k_S)$ appears in $\hat v_r$.
Since the momentum $q_r$ is also small in a near-void, but the radial velocity
$v_r$ is not, we expect $\hat v_r$ to correlate more strongly with $q_r$ 
than with $v_r$.

Since $q_r$ has the highest correlation with $\hv_r$, and is also most straightforward
to derive from an $N$-body simulation, we will use $q_r$ throughout the paper.
(In hindsight, it would make sense to rename the quadratic estimator
$\hv_r \rightarrow \hat q_r$, and call it
``kSZ momentum reconstruction'' instead of ``kSZ velocity reconstruction''. However, we
will use the $\hv_r$ notation and velocity reconstruction terminology, for consistency 
with previous papers.)
With this motivation for introducing the radial momentum, it is straightforward to compute $q_r(\x)$
from an $N$-body simulation. We simply CIC-grid particles as before, weighting each particle
by its radial velocity.

A technical point: we use compensated CIC-gridding with 1024$^3$ pixels throughout our pipeline,
even though this suppresses power at wavenumbers close
to the Nyquist frequency of the pixelization~\cite{1981csup.book.....H, Sefusatti:2015aex, Hand:2017pqn}. 
The suppression is a 3\% at $k=0.8 k_{\rm Nyq}$, and 30\% at
$k=k_{\rm Nyq}$~\cite{Hand:2017pqn}.
In our kSZ velocity reconstruction pipeline, this does not lead to biases,
provided that CIC-gridded fields and power spectra are used self-consistently
throughout the pipeline.
For example, we find (Figure~\ref{fig:quijote_correlation_and_bias} below) that the kSZ
velocity reconstruction bias $b_v$ is 1 on large scales, if the quadratic estimator
$\hv_r$ is implemented with CIC-gridding, and defined self-consistently using CIC-gridded power
spectra $P_{ge}(k_S), P_{gg}(k_S)$ (see Eq.~(\ref{eq:nvr_fourier})).

\subsection{CMB maps}
\label{ssec:pipeline_cmb}

Given the output from an $N$-body simulation, we simulate a kSZ map as follows.
Let $\x_i$ denote the 3-d position of the $i$-th particle in the simulation (where $i=1,\cdots,N_{\rm part}$),
let $\v_i$ denote the velocity, and let $\th_i = \x_{\perp i}/\chi_*$ denote the projected angular sky
location.  We approximate the momentum $\q(\x)$ as a sum of velocity-weighted delta functions:
\be
\q(\x) = \frac{1}{n_p} \sum_i \v_i \delta^3(\x-\x_i)  \label{eq:q_nbody}
\ee
where $n_p=N_{\rm part}/L_{\rm box}^3$ is the 3-d particle number density.
Plugging into the line-of-sight integral (Eq.~(\ref{eq:ksz_los})), the kSZ temperature is:
\be
T_{\rm kSZ}(\th) = \frac{K_*}{\chi_*^2 n_p} \sum_i (\hr \cdot \v_i) \delta^2(\th-\th_i)  \label{eq:ksz_sim}
\ee
In our pipeline, we discretize CMB maps using $(1024)^2$ pixels,
corresponding to angular pixel size $(\Delta\theta) = L_{\rm box}/(1024 \chi_*) = 0.96$
arcmin, and Nyquist frequency $l_{\rm Nyq} = \pi/(\Delta\theta) = 11250$.
We evaluate the RHS of Eq.~(\ref{eq:ksz_sim}) by gridding each delta function
onto the 2-d mesh using the Cloud-in-Cell (CIC) scheme~\cite{1981csup.book.....H}.

We add simulations of the lensed primary CMB and instrumental noise to our
simulations, treating both contributions as Gaussian fields.
We use noise power spectrum:
\be
N_l = s_w^2 \exp\left[ \frac{l(l+1)\theta_{\rm fwhm}^2}{8 \ln 2} \right]  \label{eq:noise_model}
\ee
with white noise level $s_w = 0.5$ $\mu$K-arcmin, and beam size $\theta_{\rm fwhm} = 1$ 
arcmin.
Note that we treat the non-kSZ CMB as Gaussian, which neglects possible
biases from non-Gaussian secondaries.
This is a loose end, although symmetry arguments suggest that biases are
probably small.
For more discussion, see~\S\ref{sec:discussion}.

In the rest of the paper, we fix fiducial survey parameters described above
($s_w=0.5$ $\mu$K-arcmin, $\theta_{\rm fwhm}=1$ arcmin, $b_g=3.24$, effective
$dn_g/dz = 0.8$ arcmin$^{-2}$).
Our galaxy survey parameters are similar to DESI, and our CMB parameters
are intentionally futuristic (a bit better than CMB-S4), in order to maximize statistical power
of our simulations.

\section{LSS and CMB power spectra}
\label{sec:power_spectrum_results}

\begin{figure}
    \centering
    \includegraphics[scale=1.
    ]{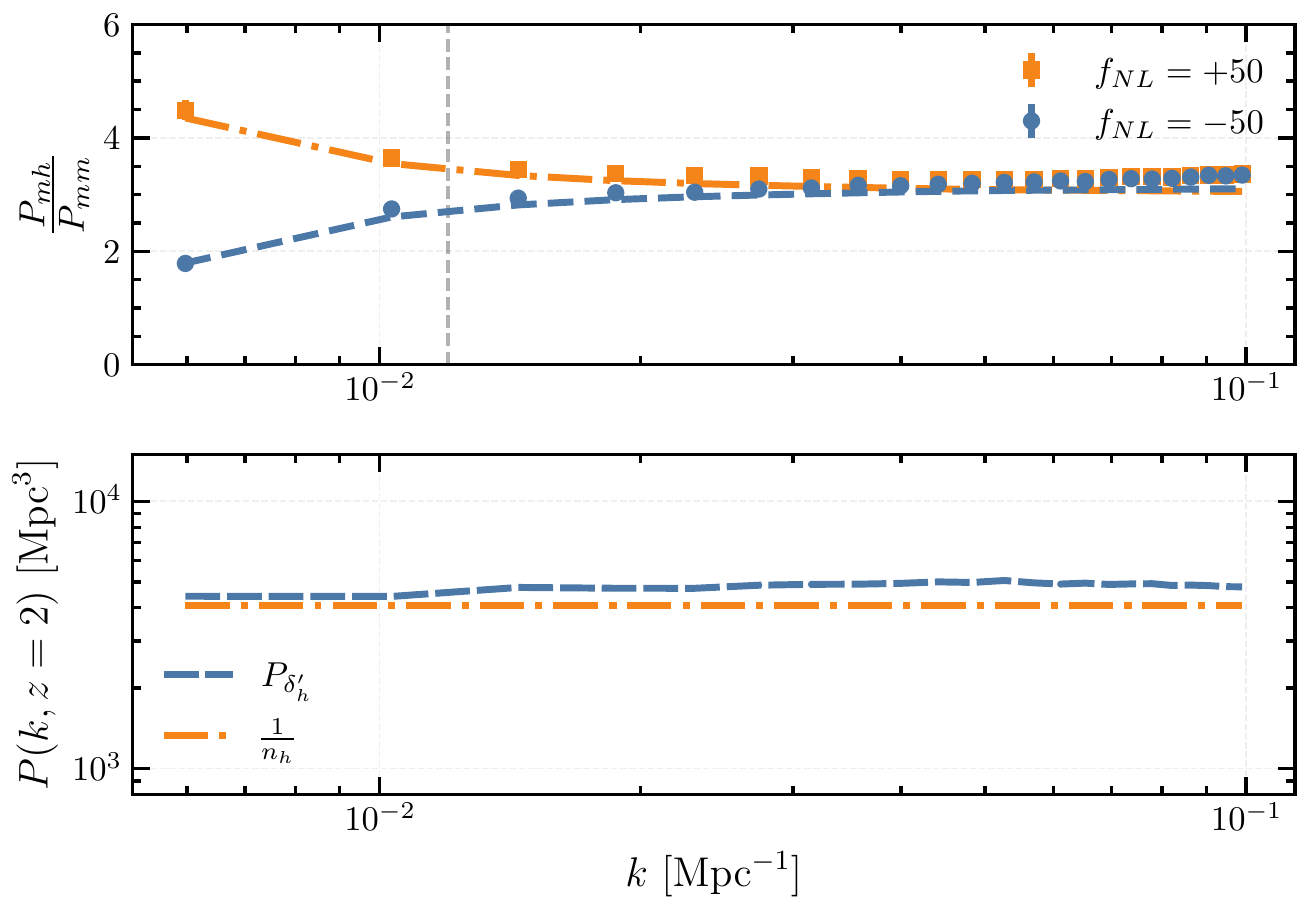}
    \caption{\emph{Top panel.} Large-scale halo bias $b_h(k) = P_{mh}(k)/P_{mm}(k)$ from $N$-body simulations
    with $f_{NL}= \pm 50$.
    For comparison, we show the bias model~(\ref{eq:bh_lowk}),
    with $b_g=3.01$,~$3.15$ for $f_{NL}=-50$,~$50$ respectively.
    (These values were obtained from the MCMC pipeline to
    be presented in~\S\ref{sec:fnl_results}.)
    For each value of $f_{NL}$, we use four $N$-body simulations with volume 1 $h^{-3}$ Gpc$^3$ each.
    \emph{Bottom panel.} 
    Halo shot noise from simulation from the same set of simulations, defined as the power spectrum of the
    field $\delta_h'(\k) = \delta_m(\k) - b_h(k) \delta_m(\k)$, compared to the Poisson
    prediction $P_{\delta_h'}(k) = 1/n_h$.}
    \label{fig:shot_noise_and_bias}
\end{figure}

In this section we present matter, halo, and CMB power spectra
from our simulation pipeline. 
In the next section we will study higher-point statistics and
kSZ velocity reconstruction.
We start by confirming that the large-scale halo bias is described by the model:
\be
b_h(k) = b_g + f_{NL} \frac{2 \delta_c(b_g-1)}{\alpha(k,z)} \hspace{1cm} (\delta_c=1.42)  \label{eq:bh_lowk}
\ee
in agreement with previous
studies~\cite{Dalal:2007cu,Desjacques_2009,Pillepich:2008ka,Grossi:2009an,Giannantonio:2009ak,Wagner:2011wx,Scoccimarro:2011pz,Biagetti:2016ywx}.
In Figure~\ref{fig:shot_noise_and_bias} (top), we estimate the halo bias
$b_h(k) = P_{mh}(k)/P_{mm}(k)$ directly from the matter-halo and matter-matter
power spectra of simulations with $f_{NL} = \pm 50$, and find good agreement 
with the bias model in Eq.~(\ref{eq:bh_lowk}).

Next we consider the halo-halo power spectrum $P_{hh}(k)$.
On large scales, we want to check that linear halo bias plus shot noise is a good description, i.e.
\be
P_{hh}(k) = b_h(k)^2 P_{mm}(k) + \frac{1}{n_h}  \label{eq:phh_lowk}
\ee
A stronger version of this check is to show that the power spectrum of the field 
$\delta_h' = \delta_h - b_h(k) \delta_m$ is consistent with pure shot noise:
$P_{\delta_h'\delta_h'}(k) = 1/n_h$. 
In Figure~\ref{fig:shot_noise_and_bias} (bottom), we find good agreement, thus
confirming the model~(\ref{eq:phh_lowk}).

\begin{figure}
    \centering
    \includegraphics[scale=0.8]{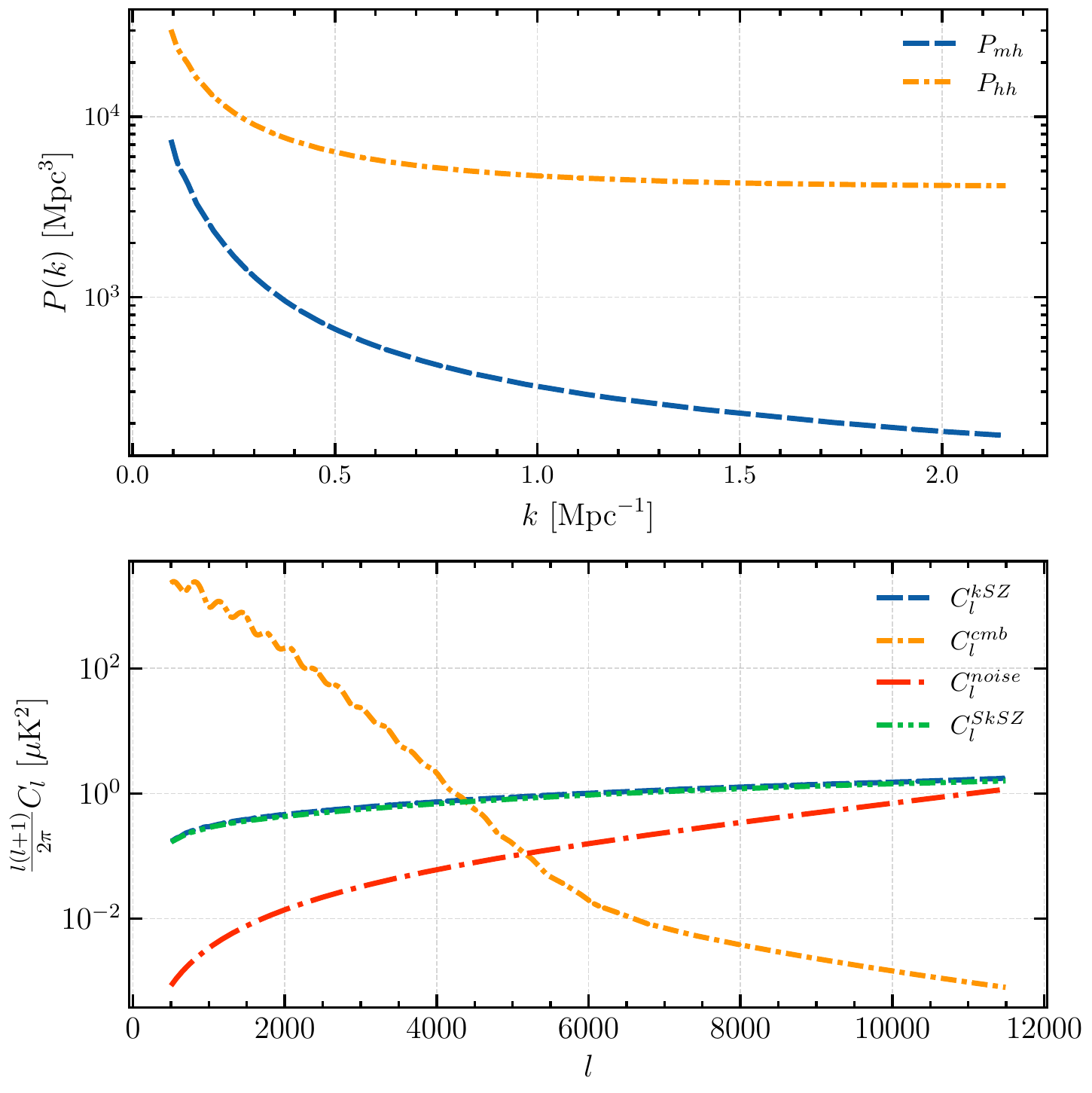}
    \caption{\emph{Top.} Small scale halo-halo and matter-halo power spectrum from Quijote N-body simulations.
    These power spectra are used in the definition of
    the velocity reconstruction estimator $\hv_r$ (Eq.~(\ref{eq:nvr_fourier}) below).
    \emph{Bottom.} Plot showing various contributions to the CMB power spectrum. The lensed CMB power spectrum is computed using the \texttt{CLASS} Boltzmann code \cite{Blas_2011}, and the noise power spectrum is based on Eq.~(\ref{eq:noise_model}) with $s_w=0.5$ $\mu$K-arcmin and $\theta_{\rm fwhm}=1$ arcmin. The kSZ power spectrum $C_l^{\rm kSZ}$ is estimated directly from our simulation pipeline. For comparison, we also show the ``standard'' analytical estimate $C_l^{SkSZ}$, based on~\cite{shaw} and computed using Eq.~(\ref{eq:analytic_ksz}).}
    \label{fig:small_scale_power_spectra}
\end{figure}

Taken together, Eqs.~(\ref{eq:bh_lowk}),~(\ref{eq:phh_lowk}) are a complete
model for halo clustering on large scales.
Turning next to small scales, we present small-scale power spectra which 
are relevant for kSZ velocity reconstruction.
The definition of the kSZ velocity reconstruction estimator $\hv_r$
(Eq.~(\ref{eq:hvr_fourier}) below) involves the small-scale galaxy-electron
and galaxy-galaxy power spectra $P_{ge}(k_S), P_{gg}(k_S)$, evaluated at
wavenumbers $k_S \sim 1$ Mpc$^{-1}$.
In the collisionless $N$-body approximation used in this paper ($\delta_e \approx \delta_m$,
$\delta_g \approx \delta_h)$, these power spectra are equal to $P_{mh}(k_S)$ and $P_{hh}(k_S)$,
which we show for reference in the top panel of Figure~\ref{fig:small_scale_power_spectra}.

The definition of $\hv_r$ also involves the small-scale CMB power spectrum $C_l^{\rm tot}$,
which is the sum of kSZ, noise, and lensed CMB contributions.
These contributions are shown in the bottom panel of Figure~\ref{fig:small_scale_power_spectra}.
The kSZ contribution $C_l^{kSZ}$ is estimated directly from the simulations.

As another check on our pipeline, in Figure~\ref{fig:small_scale_power_spectra}
we compare $C_l^{kSZ}$ to the ``standard'' analytic estimate
$C_l^{SkSZ}$, and find good agreement.
The analytic estimate is derived following~\cite{jaffe,shaw}
by approximating the electron momentum as $\q_e = (1+\delta_e) \v$
where the linear velocity $\v$ and nonlinear electron field $\delta_e$
are Gaussian.
In this approximation, the kSZ power spectrum is:
\be
C_l^{SkSZ} = \frac{(faH_{*})^2K_{*}^2 L}{2} \int
  \frac{d^3\k'}{(2\pi)^3} P^{NL}_{mm}(|\k-\k'|) P_{mm}(k') 
  \left. \frac{k(k-2k'\mu)(1-\mu^2)}{k'^2(k^2+k'^2-2kk'\mu)} 
  \right|_{\substack{\k=\l/\chi_* \\ \mu=\hat\k\cdot\hat\k'}}
  \label{eq:analytic_ksz}
\ee
where $P_{mm}$ and $P^{NL}_{mm}$ are the linear and non-linear matter power spectrum, 
and $L$ is the box size.

The kSZ power spectrum in Figure~\ref{fig:small_scale_power_spectra} underestimates
the predicted $C_l$ from hydrodynamical simulations (e.g.~\cite{Sehgal_2010})
by a factor $\sim$2.
This is because our ``snapshot'' geometry only includes kSZ fluctuations from
a redshift slice of thickness $L_{\rm box} = 1$ $h^{-1}$ Gpc.
(We also make the approximation that electrons trace dark matter, i.e.~$P_{ee} \approx P_{mm}$,
which has the opposite effect of increasing $C_l$, but this is a smaller effect.)
This is not an issue for purposes of this paper, where our goal is to test kSZ velocity 
reconstruction for biases as precisely as possible, under a self-consistent set of assumptions.
We considered making the simulations more realistic, by adding simulated kSZ outside the
simulated redshift range, but we expect that this would be nearly equivalent to adding
uncorrelated Gaussian noise, and would only serve to decrease the precision of our tests.

\section{The KSZ quadratic estimator applied to $N$-body simulations}
\label{sec:vrec_results}

\subsection{KSZ quadratic estimator}
\label{ssec:ksz_quadratic_estimator}

In this section, we describe our implementation of the kSZ
velocity reconstruction estimator $\hv_r$.
The inputs to kSZ velocity reconstruction are the 2-d CMB map $T(\l)$
and 3-d galaxy overdensity field $\delta_g(\k)$. The outputs are the
3-d radial velocity reconstruction $\hv_r(\k)$ and noise power
spectrum $N_{v_r}^{(0)}(\k_L)$. These are given by~\cite{Smith:2018bpn}:
\ba
\hv_r(\k_L) &=& N_{v_r}^{(0)}(\k_L) \frac{K_*}{\chi_*^2} \int \frac{d^3\k_S}{(2\pi)^3} \, \frac{d^2\l}{(2\pi)^2}
  \frac{P_{ge}(k_S)}{P_{gg}(k_S) C_l^{\rm tot}} \delta_g^*(\k_S) T^*(\l) \,
  (2\pi)^3 \delta^3\left(\k_L+\k_S+\frac{\l}{\chi_*}\right)  \label{eq:hvr_fourier} \\
N_{v_r}^{(0)}(\k_L) &=& \frac{\chi_*^4}{K_*^2} \left[ 
 \int \frac{d^3\k_S}{(2\pi)^3} \, \frac{d^2\l}{(2\pi)^2} \,
   \frac{P_{ge}(k_S)^2}{P_{gg}(k_S) C_l^{\rm tot}}
   (2\pi)^3 \delta^3\left( \k_L + \k_S + \frac{\l}{\chi_*} \right)
\right]^{-1}  \label{eq:nvr_fourier}
\ea
The noise power spectrum in the second line~(\ref{eq:nvr_fourier}) is obtained 
by calculating the two-point function 
$\langle \hv_r(\k_L) \hv_r(\k_L')^* \rangle$,
under the approximation that the galaxy catalog and CMB are
independent.
To emphasize an analogy with CMB lens reconstruction that will be
explained in \S\ref{sec:noise_contributions}, we will call the
noise power spectrum
defined in Eq.~(\ref{eq:nvr_fourier}) the ``kSZ $N^{(0)}$-bias''
throughout the paper. 
One of our goals is to compare the kSZ $N^{(0)}$ bias to the reconstruction
noise in $N$-body simulations, to test the accuracy of the approximation
leading to Eq.~(\ref{eq:nvr_fourier}).

In principle, $N_{v_r}^{(0)}(\k_L)$ is a function of both $k_L = |\k_L|$,
and the direction of $\k_L$ relative to the line of sight. However, on the large scales
which are relevant for constraining $f_{NL}$, it approaches a constant:
\be
N_{v_r}^{(0)}(\k_L) \rightarrow \frac{\chi_*^4}{K_*^2} \left[
 \int \frac{d^2\l}{(2\pi)^2} \frac{P_{ge}(l/\chi_*)^2}{P_{gg}(l/\chi_*) C_l^{\rm tot}}
\right]^{-1}  \hspace{1cm}  (\k_L \rightarrow 0)
\label{eq:nvr_k0}
\ee

In Eqs.~(\ref{eq:hvr_fourier}),~(\ref{eq:nvr_fourier}), we have given Fourier-space
expressions for $\hv_r(\k_L)$ and $N_{v_r}^{(0)}(\k_L)$.
These expressions are computationally expensive, and in practice alternative
expressions are used, which factorize the computation into FFT's as follows.
The velocity reconstruction $\hv_r(\k_L)$ is computed as:
\be
\hv_r(\k_L) = N_{v_r}^{(0)}(\k_L) \frac{K_*}{\chi_*^2} \int d^3\x \, \tdelta_g(\x) \, 
   \tT\Big( \frac{\x_\perp}{\chi_*} \Big) e^{-i\k_L\cdot\x} \label{eq:hvr_fast}
\ee
where the filtered galaxy field $\tdelta_g(\x)$ and filtered CMB $\tT(\th)$ are defined by:
\be
\tdelta_g(\x) = \int \frac{d^3\k_S}{(2\pi)^3} \, 
  \frac{P_{ge}(\k_S)}{P_{gg}(\k_S)} \delta_g(\k_S) e^{i\k_S\cdot\x}
\hspace{1cm}
\tT(\th) = \int \frac{d^2\l}{(2\pi)^2} \,
  \frac{1}{C_l^{\rm tot}} T(\l) e^{i\l\cdot\th}
\ee
Similarly, the kSZ $N^{(0)}$-bias is computed efficiently as:
\be
N_{v_r}^{(0)}(\k_L) = \frac{\chi_*^4}{K_*^2} \left[ 
  \int d^3\x \, f_1(\x) \, f_2\Big(\frac{\x_\perp}{\chi_*}\Big) e^{-i\k_L\cdot\x}
\right]^{-1}  \label{eq:nvr_fast}
\ee
where the 3-d field $f_1$ and 2-d field $f_2$ are defined as:
\be
f_1(\x) = \int \frac{d^3\k_S}{(2\pi)^3} \, 
\frac{P_{ge}(k_S)^2}{P_{gg}(k_S)} e^{i\k_S\cdot\x}
\hspace{1cm}
f_2(\th) = \int \frac{d^2\l}{(2\pi)^2} \, \frac{1}{C_l^{\rm tot}} \, e^{i\l\cdot\th}  \label{eq:f1f2}
\ee
Eqs.~(\ref{eq:hvr_fast}),~(\ref{eq:nvr_fast}) for $\hv_r$ and $N_{v_r}^{(0)}$
are mathematically equivalent to Eqs.~(\ref{eq:hvr_fourier}),~(\ref{eq:nvr_fourier}),
but have much lower computational cost.

One more detail of our $\hv_r$ implementation.
The definitions of $\hv_r$ and $N_{v_r}^{(0)}$ above involve small-scale power
spectra $P_{ge}(k_S)$, $P_{gg}(k_S)$, and $C_l^{\rm tot}$.
In this paper, we do not attempt to model these small-scale spectra
(e.g.~with the halo model).
Instead, we measure them directly from simulation, by estimating each power
spectrum in bandpowers, and interpolating to get a smooth function of wavenumber.
The estimated power spectra $P_{ge}(k_S)$, $P_{gg}(k_S)$, and $C_l^{\rm tot}$
in our simulations were shown previously in Figure~\ref{fig:small_scale_power_spectra}.

By estimating small-scale power spectra directly from simulation,
our pipeline is ``cheating'', since $P_{ge}(k_S)$ is not observable.
(The other two small-scale power spectra $P_{gg}(k_S)$
and $C_l^{\rm tot}$ can be estimated directly from data in a
real experiment, and so it is not cheating to measure
them from simulations.)
In a real experiment, we would need to use a fiducial model 
$P_{ge}^{\rm fid}(k_S)$, which need not equal the true power
spectrum $P_{ge}^{\rm true}(k_S)$.
In~\cite{Smith:2018bpn}, it is predicted that in this situation,
the velocity reconstruction acquires a large-scale linear bias:
\be
\hv_r(\k_L) = b_v v_r(\k_L) + (\mbox{Reconstruction noise})
\ee
where the velocity reconstruction bias $b_v$ is $1$
if $P_{ge}^{\rm fid} = P_{ge}^{\rm true}$, but can differ
from 1 if $P_{ge}^{\rm fid} \ne P_{ge}^{\rm true}$. We will
test this prediction in the next section.

\subsection{Noise and bias of velocity reconstruction}
\label{ssec:vrec_results}

In~\cite{Smith:2018bpn} we predicted that the velocity reconstruction estimator
$\hv_r(\k_L)$ is an unbiased estimator of the radial momentum $q_r(\k_L)$ on
large scales, and that the reconstruction noise is given by
the $N^{(0)}$-bias in Eq.~(\ref{eq:nvr_fourier}).
These statements are ``predictions'' since they are derived using analytic
approximations to the statistics of large-scale structure on nonlinear scales.
In this section, we will test these key predictions with $N$-body simulations.

We start by stating precisely the predictions we would like to test.
We define the kSZ velocity reconstruction bias $b_v(\k_L)$ of the simulation by:
\be
b_v(\k_L) = \frac{P_{q_r\hv_r}(\k_L)}{P_{q_rq_r}(\k_L)}
\ee
Then we predict that $b_v \rightarrow 1$ on large scales, if we
assume that the galaxy-electron power spectrum $P_{ge}(k_S)$ is known in advance
and used in the quadratic estimator~(\ref{eq:hvr_fourier}).
If fiducial power spectrum $P_{ge}^{\rm fid}(k_S) \ne P_{ge}^{\rm true}(k_S)$
is used, then we make the weaker prediction that $b_v$ approaches a constant on
large scales.

We define the reconstruction noise field $\eta(\k_L) = \hv_r(\k_L) - b_v(\k_L) q_r(\k_L)$,
or equivalently:
\be
\hv_r(\k_L) = b_v(\k_L) q_r(\k_L) + \eta(\k_L)
  \hspace{1cm}
  \mbox{where } P_{\eta q_r}(\k_L) = 0  \label{eq:eta_def}
\ee
Then we predict that the power spectrum $P_\eta(\k_L)$ is equal to the kSZ
$N^{(0)}$-bias $N_{v_r}^{(0)}(\k_L)$ given previously in Eq.~(\ref{eq:nvr_fourier}).

Note that in the above, we compare $\hv_r$ to the
radial momentum $q_r$, since $\hv_r$ is expected to be more correlated
with momentum than with other definitions of the radial velocity, and momentum
is also more straightforward to define in simulation (see discussion in~\S\ref{ssec:lss_fields}).

In the rest of this section, we will test the above predictions with simulations.
All results in this section use 100 Quijote simulations with $f_{NL}=0$ and total
volume 100 $h^{-3}$ Gpc$^3$.

\begin{figure}
    \includegraphics[scale=0.7]{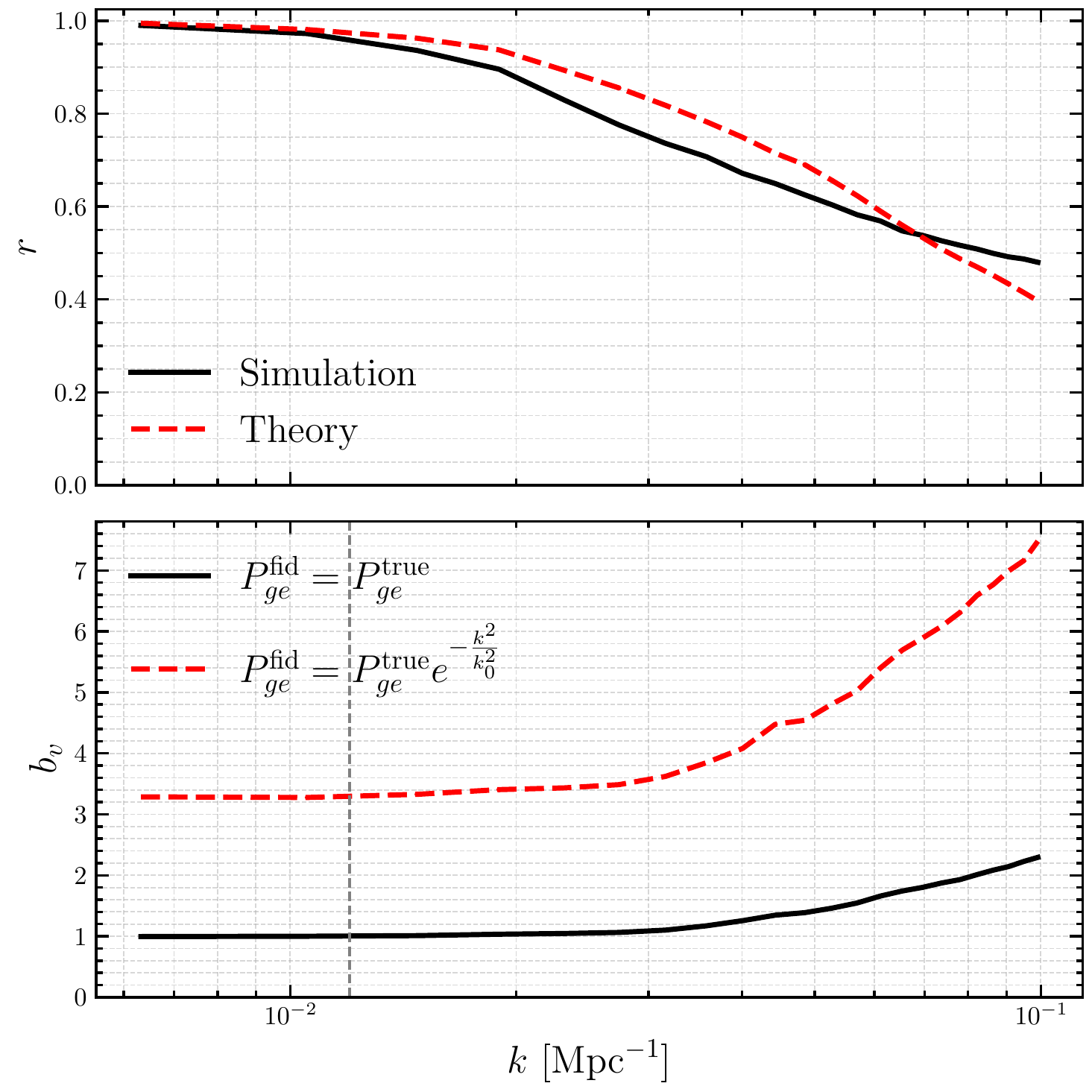}
    \caption{{\em Top panel.} Correlation coefficient $r$ between 
     fields $\hv_r$ and $q_r$, where $\hv_r$ is the kSZ velocity reconstruction
     derived from an $N$-body simulation, and $q_r$ is the true radial 
     momentum of the simulation.
     We estimate $r$ in $k$-bins using Eq.~(\ref{eq:rb_def}),
     excluding wavenumbers with $\mu=0$.
     The ``theory'' curve was obtained using Eq.~(\ref{eq:rb_theory}).
    {\em Bottom panel.} KSZ velocity reconstruction bias $b_v$, estimated in $k$-bins
     using Eq.~(\ref{eq:bvb_def}).
     The solid line was computed assuming perfect knowledge of the galaxy-electron
     power spectrum $P_{ge}(k_S)$ in the definition of $\hv_r$.
     The dashed line was computed using fiducial galaxy-electron power spectrum
     $P_{ge}^{\rm fid}(k_S) = P_{ge}^{\rm true}(k_S) \exp(-k^2/k_0^2)$,
      where $k_0=1$ Mpc$^{-1}$.
     The vertical line at $k=0.012$ Mpc$^{-1}$ is 
     the $k_{max}$ that we use in our MCMC pipeline later (\S\ref{sec:fnl_results}).}
    \label{fig:quijote_correlation_and_bias}
\end{figure}

Before exploring bias and reconstruction noise, we do a simple intuitive
comparison between the radial momentum $q_r$ and the reconstruction $\hv_r$.
In Figure~\ref{fig:quijote_correlation_and_bias} (top)
we show the correlation coefficient between $q_r$ and $\hv_r$.
More precisely, we choose a set of $k$-bins, and for each $k$-bin $b$ we define a
correlation coefficient $r_b^{\rm sim}$ by:
\be
r_b^{\rm sim} = \frac{
 \sum_{\k\in b} \hv_r^*(\k) q_r(\k)
}{\left( \sum_{\k\in b} |q_r(\k)|^2 \right)^{1/2} 
 \left( \sum_{\k\in b} |\hv_r(\k)|^2 \right)^{1/2}} \label{eq:rb_def}
\ee
It is seen that the kSZ-derived velocity reconstruction
$\hv_r(\k)$ is nearly 100\% correlated to the true momentum on large scales.
This is crucial, since we want to use velocity reconstruction to cancel sample
variance in the galaxy field and constrain $f_{NL}$, which requires a high
correlation.
To quantify this better, we compare to the ``theory'' prediction for the correlation
coefficient:
\be
r_b^{\rm theory} = \left(
\frac{\sum_{\k\in b} \mu_{\k}^2 P_{vv}(k)}{\sum_{\k\in b} (\mu_{\k}^2 P_{vv}(k) + N_{v_r}(\k))}
\right)^{1/2}
  \label{eq:rb_theory}
\ee
where $\mu_{\k} = (\hk \cdot \hr)$ as usual.
This expression for $r_b^{\rm theory}$ was calculated assuming $b_v=1$ and
$P_\eta=N_{v_r}^{(0)}$.

In Figure~\ref{fig:quijote_correlation_and_bias} (top), the correlation coefficient
seen in simulation qualitatively agrees with the theory prediction, but we do see some level
of mismatch.
On large scales, $r_b^{\rm sim}$ is a little smaller than $r_b^{\rm theory}$.
This is consistent with a factor 2--3 increase in reconstruction noise that we
will describe shortly.
Intriguingly, on small scales, $r_b^{\rm sim}$ is a little larger than $r_b^{\rm theory}$.
By comparing Figs~\ref{fig:quijote_correlation_and_bias} (bottom) and~{\ref{fig:quijote_mean_powerspectrum},
this can be interpreted as arising from enhancement of the velocity bias $b_v$ on small scales,
with no corresponding enhancement in reconstruction noise. 

Next we would like to test the prediction that the velocity reconstruction bias
$b_v \rightarrow 1$ on large scales.
In Figure~\ref{fig:quijote_correlation_and_bias} (bottom), we show the
bias from $N$-body simulations, estimated in non-overlapping $k$-bins
by defining:
\be
(b_v)_b = \frac{\sum_{\k\in b} q_r(\k)^* \, \hv_r(\k)}{\sum_{\k\in b} |q_r(\k)|^2}  \label{eq:bvb_def}
\ee
for each $k$-bin $b$.
The bias is 1 on large scales as predicted.
As $k$ increases, the bias is an increasing function of $k$, and becomes large
for surprisingly small values of $k$.
For example, $b_v \approx 2.4$ at $k=0.1$ Mpc$^{-1}$.
The level of scale dependence seen in the velocity bias $b_v(k)$ is much higher than
the familiar case of halo bias (see Fig.~\ref{fig:shot_noise_and_bias}).
However, on the very large scales $(k \lesssim 0.01$) that are important for
$f_{NL}$ constraints, $b_v$ is constant to an excellent approximation.

In Figure~\ref{fig:quijote_correlation_and_bias} (bottom), we also show
the velocity bias $b_v$ if we construct the quadratic estimator $\hv_r$
using fiducial galaxy-electron power spectrum 
$P_{ge}^{\rm fid}(k_S) \ne P_{ge}^{\rm true}(k_S)$.
For illustrative purposes, we have arbitrary chosen
$P_{ge}^{\rm fid}(k_S) = P_{ge}^{\rm true}(k_S) \exp(-k^2/k_0^2)$, where
$k_0 = 1$ Mpc$^{-1}$.
As predicted, we find that $b_v$ approaches a constant on large scales,
but the value is $\ne 1$.

\begin{figure}
    \includegraphics[]{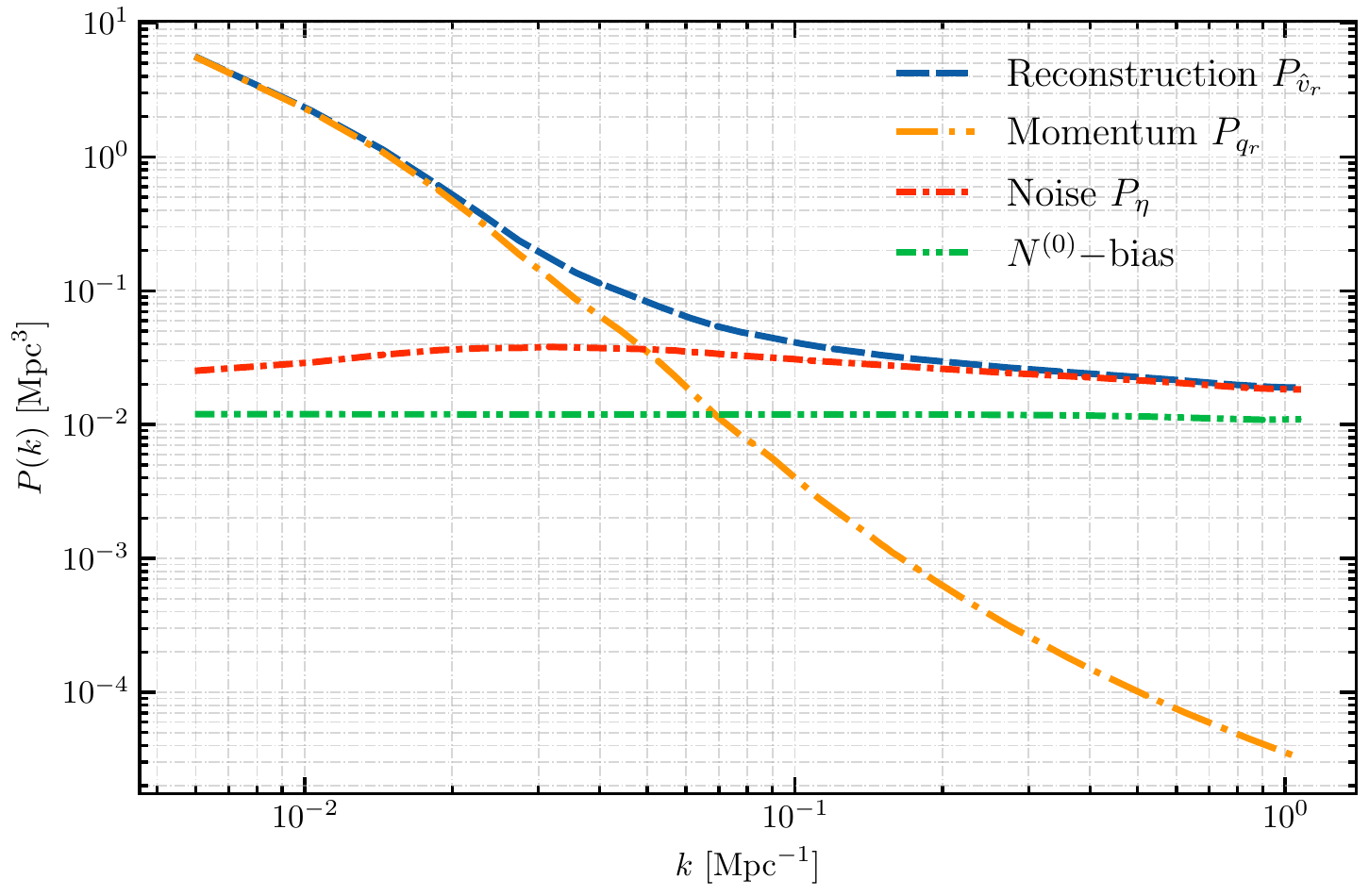}
    \caption{Velocity reconstruction signal and noise power spectra from 
    100 high resolution Quijote simulations.
    $P_{\hv_r}$ and $P_{q_r}$ are the power spectra of the noisy
    velocity reconstruction $\hv_r$, and the true radial momentum $q_r$
    (both from simulation). 
    $P_\eta$ is the reconstruction noise from simulation, defined
    as the power spectrum of the noise field $\eta$ defined in Eq.~(\ref{eq:eta_def}).
    $N_{v_r}^{(0)}$ is the kSZ $N^{(0)}$-bias in Eq.~(\ref{eq:nvr_fourier}).}
    \label{fig:quijote_mean_powerspectrum}
\end{figure}

Finally, we come to the main result of this section:
comparing the reconstruction noise $P_\eta$
in simulation with the kSZ $N^{(0)}$-bias.
In Figure~\ref{fig:quijote_mean_powerspectrum}, we show four power
spectra:
\begin{itemize}
    \item The total power spectrum $P_{\hv_r}$ of the kSZ velocity
    reconstruction (including noise), estimated from simulation.
    \item The power spectrum $P_{q_r}$ of the radial momentum,
    estimated from simulation.
    \item The reconstruction noise power spectrum $P_{\eta}$,
    estimated from simulation using the definition
    of $\eta$ in Eq.~(\ref{eq:eta_def}).
    \item The kSZ $N^{(0)}$-bias $N_{v_r}^{(0)}$, computed using
    Eq.~(\ref{eq:nvr_fourier}). 
\end{itemize}
Contrary to the prediction from~\cite{Smith:2018bpn},
the reconstruction noise $P_\eta$ in simulation
exceeds the kSZ $N^{(0)}$-bias by a factor 2--3!
This increase in noise can potentially affect $f_{NL}$ constraints,
even though the $f_{NL}$ constraints are derived from large scales
where the velocity reconstruction is signal-dominated, because sample
variance cancellation plays a role in the constraints.
We will explore this issue in more detail in~\S\ref{sec:noise_contributions}
and~\S\ref{sec:fnl_results}.

As a code check, we also estimated the power spectrum of a 
``fake'' kSZ velocity reconstruction $\hv_r^{\rm fake}$,
constructed by applying the quadratic estimator to a
galaxy catalog $\delta_g$ and a CMB map $T$ derived from
independent $N$-body simulations.
The power spectrum of $\hv_r^{\rm fake}$ is exactly equal to 
$N_{v_r}^{(0)}$, since by construction $N_{v_r}^{(0)}$ is the
reconstruction noise under the approximation that $\delta_g$
and $T$ are independent.
In our simulations, we find the expected exact agreement
between $P_{\eta}^{\rm fake}$ and $N_{v_r}^{(0)}$.
This is a strong check on our pipeline, and indicates that
the discrepancy between $P_\eta$ and $N_{v_r}^{(0)}$
is a real effect arising from higher-point correlations in
the $N$-body simulation. In~\S\ref{sec:noise_contributions},
we will explain this discrepancy using the halo model.

\subsection{Bandpower covariance}
\label{ssec:bandpower_covariance}

So far, our comparisons between theory and simulation
have focused on mean power spectra: either the cross
spectrum $P_{\hv_r q_r}$ which determines the bias $b_v(k)$,
or the noise power spectrum $P_{\eta\eta}$.
However, for either forecasts or data analysis, the power
spectrum {\em covariance} is also important.
If the reconstruction noise $\eta$ were a Gaussian field,
then its power spectrum covariance would be:
\be
\mbox{Cov}(P_\eta(b), P_\eta(b')) = \frac{2}{N_b} P_\eta(b)^2 \delta_{bb'}
 \label{eq:gaussian_bandpower_covariance}
\ee
where $b,b'$ denote narrow non-overlapping $k$-bins, and $N_b$
denotes the number of modes in bin $b$.
The standard Fisher matrix forecasting formalism implicitly
assumes that the Gaussian bandpower
covariance~(\ref{eq:gaussian_bandpower_covariance})
is a good approximation.
Our MCMC $f_{NL}$ pipeline in~\S\ref{sec:fnl_results}
will make slightly stronger assumptions, by assuming that
the full probability density function of $\eta$ is well-described
by its Gaussian approximation.

As one test of the Gaussian approximation, we estimate the
correlation matrix between bandpowers and show the result
in Figure~\ref{fig:correlation3}.
We find that non-Gaussian bandpower covariance is small at 
low $k$, but very significant (correlations of order one)
at high $k$.
The transition between the two regimes is fairly sharp and
occurs at $k \sim 0.03$ Mpc$^{-1}$. 
This suggests that non-Gaussian bandpower covariance is
unlikely to be an issue for constraining $f_{NL}$, where statistical
weight comes from the very largest scales.
(For example, in the $f_{NL}$ analysis in the next 
section, we will use $k_{\rm max} = 0.012$ Mpc$^{-1}$.)
However, the bandpower covariance in
Figure~\ref{fig:correlation3} assumes our fiducial survey
parameters, and we have not explored parameter dependence
systematically.

\begin{figure}
    \includegraphics[scale=0.75]{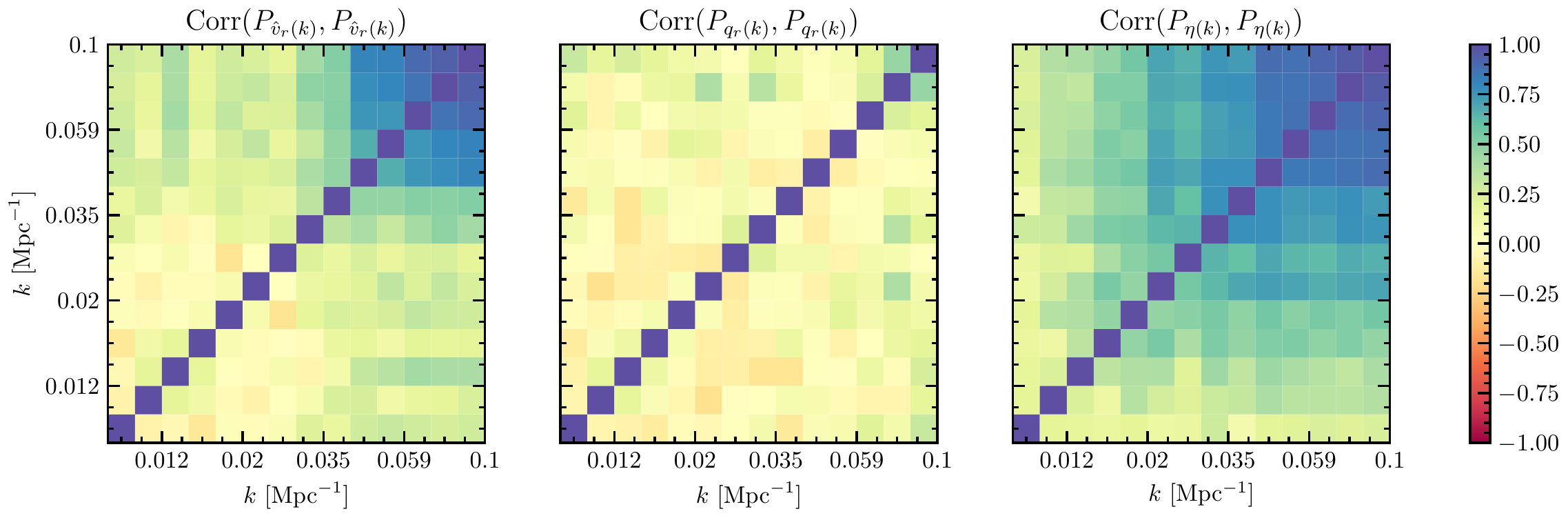}
    \caption{Correlation coefficient between bandpowers
    $P(k)$ of the velocity reconstruction $\hv_r(\k)$
    (left panel), true momentum $q_r(\k)$ (middle panel),
    and reconstruction noise $\eta(k) = \hv_r(\k) - b_v(\k) q_r(\k)$
    (right panel). Correlation coefficients were estimated
    from 100 Quijote simulations.}
    \label{fig:correlation3}
\end{figure}

\section{Higher-order biases to kSZ reconstruction noise}
\label{sec:noise_contributions}

In~\S\ref{ssec:vrec_results}, we found a discrepancy
between the reconstruction
noise $P_\eta$ in simulation and the kSZ $N^{(0)}$-bias.
In this section, we will elaborate on our previous statements that the
$N^{(0)}$-bias does not include all terms in the reconstruction noise,
derive additional terms which arise in the halo model, and numerically
compare the new terms to the simulations.

\subsection{Setup}

We will calculate the total power spectrum of the reconstruction
$P_{\hv_r\hv_r}(\k_L)$, which will contain all signal and noise
terms. First, we set up the calculation using schematic notation
which just keeps track of how many terms are present, and how each
term factorizes as a product of fields. Since the quadratic estimator
has schematic form $\hv_r \sim (\delta_g T)$, its power spectrum has
schematic form:
\be
P_{\hv_r\hv_r}(k_L) \sim \big\langle (\delta_g T) (\delta_g T) \big\rangle \, .
\ee
Our calculation will make a series of approximations which we will explain as we
go along. First, we make the approximations:
\begin{itemize}
\item {\em Approximation 1.}
We write the CMB as $T = T_{kSZ} + T_{\rm other}$,
and make the approximation that the non-kSZ contribution
$T_{\rm other}$ is statistically independent of the galaxy catalog.
This neglects possible non-Gaussian effects from
CMB secondaries, e.g.~CMB lensing.
\item {\em Approximation 2.} The electron radial momentum factorizes as $q_r = (1+v_r) \delta_e$
 in the kSZ line-of-sight integral~(\ref{eq:ksz_los}).
\end{itemize}
Under these approximations, we can write $P_{\hv_r\hv_r}(\k_L)$ schematically as:
\be
P_{\hv_r\hv_r}(\k_L) \sim
\big\langle \delta_g \delta_g \big\rangle \, 
\big\langle T_{\rm other} T_{\rm other} \big\rangle +
\big\langle (\delta_g v_r \delta_e) (\delta_g v_r \delta_e) \big\rangle
\ee
We write the six-point function
$\langle (\delta_g v_r \delta_e) (\delta_g v_r \delta_e) \rangle$
appearing on the RHS as a sum over Wick
contractions, plus a non-Gaussian part
$\langle (\delta_g v_r \delta_e) (\delta_g v_r \delta_e) \rangle_{ng}$.
There are 15 Wick contractions, but we make the following
approximation, which reduces the number to 3:
\begin{itemize}
\item {\em Approximation 3.} In the Gaussian part of the six-point
  function $\langle (\delta_g v_r \delta_e) (\delta_g v_r \delta_e) \rangle$,
  terms where $v_r$ Wick-contracts with either $\delta_g$ or $\delta_e$
  are negligible.

  The rationale for this approximation is as follows.
  The kSZ velocity reconstruction $\hv_r(\k_L)$ is determined
  by the galaxy and electron fields $\delta_g(\k), \delta_e(\k)$
  on ``kSZ'' scales $k \sim 1$ Mpc$^{-1}$.
  On these scales, radial velocity modes $v_r(\k)$ are very small,
  which implies that terms proportional to $P_{gv_r}(k)$ and
  $P_{ev_r}(k)$ should also be small.
\end{itemize}
In this approximation, $P_{\hv_r}$ has schematic form:
\be
P_{\hv_r}(\k_L) = 
\big\langle \delta_g \delta_g \big\rangle \, 
\big\langle T_{\rm other} T_{\rm other} \big\rangle +
\wick{321}{( <1 \delta_g <2 v_r <3 \delta_e ) ( >1 \delta_g >2 v_r >3 \delta_e )} 
+ \wick{212}{( <1 \delta_g <3 v_r >1 \delta_e ) ( <2 \delta_g >3 v_r g >2 \delta_e )} 
+ \wick{321}{( <1 \delta_g <3 v_r <2 \delta_e ) ( >2 \delta_g >3 v_r >1 \delta_e )} 
+ \big\langle (\delta_g v_r \delta_e) 
   (\delta_g v_r \delta_e) \big\rangle_{ng}
\label{eq:pvr_wick_contractions}
\ee
where the non-Gaussian $n$-point function $\langle \cdot \rangle_{ng}$
denotes the expectation value after subtracting all Wick contractions.

Detailed calculation of each term now shows that the
first two terms combine to give the $N^{(0)}$-bias,
the third term is the ``signal'' power spectrum $P_{v_r}(\k_L)$, and
the fourth and fifth terms are new reconstruction noise terms
$N^{(1)}$ and $N^{(3/2)}$:
\be
P_{\hv_r\hv_r}(\k_L) = 
P_{v_r}(\k_L) + N_{v_r}^{(0)}(\k_L) + N^{(1)}(\k_L) 
+ N^{(3/2)}(\k_L) \label{eq:n0_n1}
\ee
where the precise (non-schematic) forms
of the new bias terms $N^{(1)}$ and $N^{(3/2)}$ are:
\begin{align}
N^{(1)}(\k_L) &= N_{v_r}^{(0)}(\k_L)^2 \, \frac{K_*^4}{\chi_*^8}
  \int \frac{d^2\l}{(2\pi)^2} \, \frac{d^2\l'}{(2\pi)^2} \,
  \left( 
      \frac{P_{ge}(k_S)^2}{P_{gg}(k_S) C_l^{\rm tot}}
      \frac{P_{ge}(k'_S)^2}{P_{gg}(k'_S) C_{l'}^{\rm tot}} 
      P_{v_r}(\q)
  \right)_{\substack{\k_S=\k_L+\l/\chi_* \\ \k'_S=\k_L+\l'/\chi_* \\ \q = -\k_L + (\l+\l')/\chi_*}}
  \label{eq:n1_exact} \\
N^{(3/2)}(\k_L)
 &= N_{v_r}^{(0)}(\k_L)^2 \frac{K_*^4}{\chi_*^8} 
 \int \frac{d^2\l}{(2\pi)^2} \frac{d^2\l'}{(2\pi)^2} \frac{d^3\q}{(2\pi)^3} \frac{d^3\q'}{(2\pi)^3} \nn \\
 & \hspace{2.5cm} \times
 \Bigg[
  \frac{P_{ge}(k_S)}{P_{gg}(k_S) C_l^{\rm tot}}
  \frac{P_{ge}(k_S')}{P_{gg}(k_S') C_{l'}^{\rm tot}} 
  \Big\langle \delta_g(\k_S) v_r(\q) \delta_e(\p)
    \delta_g^*(\k_S') v_r^*(\q') \delta_e^*(\p') \Big\rangle'_{ng}
    \Bigg]_{\substack{
     \k_S = \k_L - \l/\chi_* \\ 
     \k_S' = \k_L - \l'/\chi* \\
     \p = -\q + \l/\chi_* \\
     \p' = -\q' + \l'/\chi_*}}  \label{eq:n32_sixpoint}
\end{align}
Here, a primed $N$-point function 
$\langle \prod_{i=1}^N X_i(\k_i) \rangle'$
denotes the expectation value without the 
delta function $(2\pi)^3 \delta^3(\sum\k_i)$.

We have chosen to call the new terms the
kSZ $N^{(1)}$-bias and $N^{(3/2)}$-bias,
to emphasize an analogy with CMB lensing.
The $N^{(0)}$ and $N^{(1)}$ biases represent the total 
KSZ reconstruction noise if all LSS fields are
Gaussian. The $N^{(1)}$-bias is a Wick contraction
which is more difficult to compute, since the integrals
cannot be factored into a sequence of convolutions.
The $N^{(3/2)}$-bias represents additional noise bias
arising from non-Gaussianity of the LSS fields
($\delta_g$ and $\delta_e$).
All of these statements are also true for the
CMB lensing $N^{(1)}$-bias~\cite{Kesden:2003cc}
and $N^{(3/2)}$-bias~\cite{Bohm:2016gzt}.
However, the analogy is not perfect: in the CMB
lensing case, there is a systematic expansion in
powers of the lensing potential $\phi$, and there
is no analogous expansion in the kSZ case. On a related
note, when we evaluate the kSZ biases
numerically, we will find that the $N^{(1)}$-bias
is much smaller than $N^{(0)}$, whereas the $N^{(3/2)}$-bias
is comparable to $N^{(0)}$.

\subsection{KSZ $N^{(1)}$-bias}

In the limit $k_L \ll k_S$, the $N^{(1)}$-bias in Eq.~(\ref{eq:n1_exact})
can be simplified a lot.
We make the following approximations inside the integral:
\be
N_{v_r}^{(0)}(\k_L) \approx N_{v_r}^{(0)}(0)
  \hspace{1cm}
\k_S = \k_L + \l/\chi_* \approx \l/\chi_*
  \hspace{1cm}
\k_S' = -\q + \l/\chi_* \approx \l/\chi_*  \label{eq:n1_lowk_approximations}
\ee
where the third approximation is valid since the integrand contains the 
factor $P_{v_r}(\q)$, which peaks for $q \ll k_S$.
Making these approximations in Eq.~(\ref{eq:n1_exact}), and changing variables
from $\l'$ to $\l'' = (\l+\l')$, the integral factorizes as:
\be
N^{(1)}(\k_L) \approx N_{v_r}^{(0)}(0)^2 \, \frac{K_*^4}{\chi_*^8}
\left( \int \frac{d^2\l}{(2\pi)^2} \, 
 \frac{P_{ge}(k_S)^4}{P_{gg}(k_S)^2 (C_l^{\rm tot})^2}
\right)_{k_S=l/\chi_*}
\left( \int \frac{d^2\l''}{(2\pi)^2} \, P_{v_r}(\q) 
\right)_{\q=-\k_L+\l''/\chi_*}  \label{eq:n1_lowk1}
\ee
We simplify the second factor as:
\begin{align}
\left( \int \frac{d^2\l''}{(2\pi)^2} \, P_{v_r}(\q) \right)_{\q=-\k_L+\l''/\chi_*}
  &= k_{Lr}^2 \int \frac{d^2\l''}{(2\pi)^2} \,
     \left( \frac{P_v(q)}{q^2} \right)_{\q=-\k_L+\l''/\chi_*}
  & \mbox{since $P_{v_r}(q) = (k_{Lr}/q)^2 P_v(q)$} \nn \\
  &= \frac{k_{Lr}^2 \chi_*^2}{2\pi}  \int_{|k_{Lr}|}^\infty dq \frac{P_v(q)}{q}  & \mbox{by change of variables}
\end{align}
To make the first factor more intuitive, we define the dimensionless quantity:
\be
W(l) = N_{v_r}^{(0)}(0) \frac{K_*^2}{\chi_*^4} 
  \left( \frac{P_{ge}(k_S)^2}{P_{gg}(k_S) 
  C_l^{\rm tot}} \right)_{\k_S=\l/\chi_*}  \label{eq:Wl_def}
\ee
which satisfies (using Eq.~(\ref{eq:nvr_fourier})):
\be
\int \frac{d^2\l}{(2\pi)^2} \, W(l) = 1 \, .
\ee
Plugging into Eq.~(\ref{eq:n1_lowk1}) we get:
\be
N^{(1)}(\k_L) \approx \frac{k_{Lr}^2 \chi_*^2}{2\pi} 
  \left( \int \frac{d^2\l}{(2\pi)^2} \, W(l)^2 \right) 
  \left( \int_{|k_{Lr}|}^\infty dq \frac{P_v(q)}{q} \right)
  \hspace{1.5cm} (k_L \ll k_S)  \label{eq:n1_lowk}
\ee
Note that for $k_L \ll k_S$, the $N^{(1)}$-bias only depends on 
$|k_{Lr}| = |\mu| k_L$.
In the limit $k_{Lr} \ll k_{eq}$, where $k_{eq} \sim 0.02$ Mpc$^{-1}$
is the matter-radiation equality scale, $N^{(1)}(k_{Lr})$ is
proportional to $|k_{Lr}|$. (In contrast to the $N^{(0)}$-bias, which
is constant on large scales.)

It will also be useful to have an expression for the $N^{(1)}$-bias
after angle-averaging $\k_L$ (e.g.~in Figure~\ref{fig:N3/2 bias}
below).
We omit the details of the calculation and quote the final result:
\be
N^{(1)}(k_L)_{\rm avg} \approx \frac{\chi_*^2}{6\pi}
 \left( \int \frac{d^2\l}{(2\pi)^2} \, W(l)^2 \right) 
 \left(
   \frac{1}{k_L} \int_0^{k_L} dq \, q^2 P_v(q)
   + k_L^2 \int_{k_L}^\infty dq \, \frac{P_v(q)}{q}
  \right) \label{eq:n1_avg}
\ee
where ``avg'' means ``angle-averaged over $\k_L$'',
and $k_L \ll k_S$ has been assumed.

\subsection{KSZ $N^{(3/2)}$-bias and halo model evaluation}
\label{ssec:n32}

In the limit $k_L \ll k_S$, the $N^{(3/2)}$-bias in 
Eq.~(\ref{eq:n32_sixpoint}) also simplifies.
We start by using the halo model to compute the non-Gaussian 
six-point function:
\be
\big\langle 
\delta_g(\k_1) \delta_e(\k_2) 
\delta_g(\k_3) \delta_e(\k_4)
v_r(\k_5) v_r(\k_6)
\big\rangle'_{ng}
\label{eq:sixpt_ng}
\ee
which appears in $N^{(3/2)}$.

We briefly summarize the ingredients of the halo model; for a
systematic review see~\cite{Cooray:2002dia}.
Let $n(M)$ be the halo mass function, or number of halos per
unit volume per unit halo mass.
Let $b(M)$ be the large-scale bias of a halo of mass $M$.
Let $n_h = \int_{M_{\rm min}}^\infty dM \, n(M)$ be the mean halo number density,
and let $\rho_m$ be the mean matter density.
Here, $M_{\rm min}$ is the minimum halo mass for our catalog (corresponding
to 40 particles).
Let $u_M(k)$ be the Fourier-transformed mass profile of a halo of mass $M$,
normalized so that $u_M(0)=1$.

It will be convenient to define:
\begin{align}
    \alpha_n(k_1,\cdots,k_n) &=
      \frac{1}{n_h} \int_{M_{\rm min}}^\infty dM \, n(M) \, 
      \prod_{i=1}^n \frac{M u_M(k_i)}{\rho_m} 
      \label{eq:alpha_def} \\
    \beta_n(k_1,\cdots,k_n) &=
      \frac{1}{n_h} \int_{M_{\rm min}}^\infty dM \, n(M) b(M) \,
      \prod_{i=1}^n \frac{M u_M(k_i)}{\rho_m}
      \label{eq:beta_def} \\
    \beta_n'(k_1,\cdots,k_n) &=
      \frac{1}{n_h} \int_{0}^\infty dM \, n(M) b(M) \,
      \prod_{i=1}^n \frac{M u_M(k_i)}{\rho_m}
      \label{eq:beta_prime_def}
\end{align}
Note that for $n=0$, we have $\alpha_0=1$ and $\beta_0=b$, where
$b = n_h^{-1} \int_{M_{\rm min}}^\infty dM \, n(M) b(M)$ is the halo bias.

Under the assumptions of the halo model, the connected six-point
function in Eq.~(\ref{eq:sixpt_ng}) can be calculated exactly.
In Appendix~\ref{app:diagrams}, we present the details of the calculation,
and diagrammatic rules for calculating $n$-point functions in the halo
model, which may be of more general interest.
In the next few paragraphs 
(Eqs.~(\ref{eq:sixpt_factorization})--(\ref{eq:S4})),
we summarize the result of the calculation.

We assume that the radial velocity modes $v_r(\k_5), v_r(\k_6)$
in the six-point function~(\ref{eq:sixpt_ng}) are evaluated on
linear scales $\k_5,\k_6$.
Then the six-point function factorizes into
lower-order correlation functions (i.e.~there are no fully
connected contributions).
More precisely, the six-point function~(\ref{eq:sixpt_ng})
is given by:
\begin{eqnarray}
&& \big\langle 
\delta_g(\k_1) \delta_e(\k_2) 
\delta_g(\k_3) \delta_e(\k_4)
v_r(\k_5) v_r(\k_6)
\big\rangle_{ng} \nn \\
&& \hspace{1cm}
= \Big[ \big( Q^{ge}_{\k_1\k_2\k_5} Q^{ge}_{\k_3\k_4\k_6} 
+ Q^{gg}_{\k_1\k_3\k_5} Q^{ee}_{\k_2\k_4\k_6}
+ Q^{ge}_{\k_1\k_4\k_5} Q^{ge}_{\k_2\k_3\k_6} \big)
+ \big(\k_5 \leftrightarrow \k_6 \big) \Big] \nn \\
&& \hspace{1.5cm}
+ \Big[ \big(
  P^{\delta_g v_r}_{\k_1\k_5} R^{ege}_{\k_2\k_3\k_4\k_6} +
  P^{\delta_e v_r}_{\k_2\k_5} R^{gge}_{\k_1\k_3\k_4\k_6} +
  P^{\delta_e v_r}_{\k_3\k_5} R^{gee}_{\k_1\k_2\k_4\k_6} +
  P^{\delta_e v_r}_{\k_4\k_5} R^{geg}_{\k_1\k_2\k_3\k_6}
  \big) + \big(\k_5 \leftrightarrow \k_6 \big) \Big] \nn \\
&& \hspace{1.5cm}
+ \Big[ S_{\k_1\k_2\k_3\k_4} P^{v_rv_r}_{\k_5\k_6} \Big]
\label{eq:sixpt_factorization}
\end{eqnarray}
where we have introduced the following notation for
some 2-, 3-, and 4-point functions:
\begin{align}
P^{XY}_{\k_1\k_2} 
  &= \big\langle X(\k_1) Y(\k_2) \big\rangle
  & (X,Y \in \{\delta_g,\delta_e,v_r\}) \\
Q^{XY}_{\k_1\k_2\k_3}
  &= \big\langle \delta_X(\k_1) \delta_Y(\k_2) v_r(\k_3) \big\rangle
  & (X,Y \in \{g,e\}) \\
R^{XYZ}_{\k_1\k_2\k_3\k_4} 
  &= \big\langle \delta_X(\k_1) \delta_Y(\k_2) \delta_Z(\k_3) v_r(\k_4) \big\rangle_{ng}
  & (X,Y,Z \in \{g,e\}) \\
S_{\k_1\k_2\k_3\k_4} 
  &= \big\langle \delta_g(\k_1) \delta_e(\k_2) \delta_g(\k_3) \delta_e(\k_4) \big\rangle_{ng}
\end{align}
The quantities $Q,R,S$ on the RHS of Eq.~(\ref{eq:sixpt_factorization})
are given explicitly by:
\begin{align}
Q^{ge}_{\k_1\k_2\k_3} &= \beta_1(k_2)
\left( \frac{ik_{3r}}{k_3} \right)
P_{mv}(k_3) \, (2\pi)^3 \delta^3\left( \smallsum\k_i \right) \\
Q^{gg}_{\k_1\k_2\k_3} &= \frac{b}{n_h}
\left( \frac{ik_{3r}}{k_3} \right)
P_{mv}(k_3) \, (2\pi)^3 \delta^3\left( \smallsum\k_i \right) \\
Q^{ee}_{\k_1\k_2\k_3} &= n_h \beta_2'(k_1,k_2)
\left( \frac{ik_{3r}}{k_3} \right)
P_{mv}(k_3) \, (2\pi)^3 \delta^3\left( \smallsum\k_i \right) \\
R^{gge}_{\k_1\k_2\k_3\k_4} &= \frac{\beta_1(k_3)}{n_h}
\left( \frac{ik_{4r}}{k_4} \right)
P_{mv}(k_4) \, (2\pi)^3 \delta^3\left(\smallsum\k_i \right) \\
R^{gee}_{\k_1\k_2\k_3\k_4} &= \beta_2(k_2,k_3)
\left( \frac{ik_{4r}}{k_4} \right)
P_{mv}(k_4) \, (2\pi)^3 \delta^3\left(\smallsum\k_i \right) \\
S_{\k_1\k_2\k_3\k_4} &= \Bigg[ \frac{\alpha_2(k_2,k_4)}{n_h}
+ \beta_1(k_2) \beta_1(k_4) P_{\rm lin}(\k_1+\k_2)
+ b \beta_2'(k_2,k_4) P_{\rm lin}(\k_1+\k_3)
\nn \\ & \hspace{0.75cm}
+ \beta_1(k_2) \beta_1(k_4) P_{\rm lin}(\k_1+\k_4)
+ b \beta_2(k_2,k_4) P_{\rm lin}(k_1)
+ \beta_1'(k_2) \beta_1(k_4) P_{\rm lin}(k_2)
\nn \\ & \hspace{0.75cm}
+ b \beta_2(k_2,k_4) P_{\rm lin}(k_3)
+ \beta_1(k_2) \beta_1'(k_4) P_{\rm lin}(k_4) \bigg]
(2\pi)^3 \delta^3(\smallsum\k_i)
\label{eq:S4}
\end{align}
where $P_{mv}(k) = (faH/k) P_{\rm lin}(k)$ is the linear
matter-velocity power spectrum.

Taken together, Eqs.~(\ref{eq:sixpt_factorization})--(\ref{eq:S4}))
are a complete calculation of the six-point function~(\ref{eq:sixpt_ng})
in the halo model, in a rather daunting form with 22 terms! However,
we will now argue that most of these terms are negligible, when
we compute the $N^{(3/2)}$-bias by plugging the six-point function into
the integral~(\ref{eq:n32_sixpoint}).

In the integral~(\ref{eq:n32_sixpoint}), the six-point function
is evaluated at the following configuration of wavenumbers
$\k_1,\cdots,\k_6$:
\be
\k_1 = \k_L - \frac{\l}{\chi_*}
\hspace{0.8cm}
\k_2 =  -\q + \frac{\l}{\chi_*}
\hspace{0.8cm}
\k_3 = -\k_L + \frac{\l'}{\chi_*}
\hspace{0.8cm}
\k_4 = \q' - \frac{\l'}{\chi_*}
\hspace{0.8cm}
\k_5 = \q
\hspace{0.8cm}
\k_6 = -\q'
\ee
To understand which terms are negligible, we classify
wavenumbers as either ``small-scale'' (meaning a typical
kSZ scale $\sim$1 Mpc$^{-1}$), or ``large-scale''
(meaning $\ll 1$ Mpc$^{-1}$).
In the integral~(\ref{eq:n32_sixpoint}), we formally integrate
over all wavenumbers $(\l,\l',\q,\q')$, but we will assume that
$q,q'$ are large-scale , and $(l/\chi_*),(l'/\chi_*)$ are small-scale,
since these wavenumber configurations dominate the integral.
We will also assume that $k_L$ is large-scale,
since we are interested in the $N^{(3/2)}$-bias in the limit
$k_L \rightarrow 0$.

Now we can state our criteria for deciding which terms
in the six-point function are negligible:

\begin{itemize}
    \item {\em Approximation 4.} In the six-point 
    function~(\ref{eq:sixpt_ng}), terms where
    $P_{\rm lin}$ is evaluated at
    a small-scale wavenumber give negligible
    contributions to $N^{(3/2)}$. 
    
    Rationale: On a small scale $k$, clustering is small compared to
    halo shot noise, so terms in the reconstruction noise
    proportional to $P_{\rm lin}(k)$ should be
    subdominant to other contributions.
    
    \item {\em Approximation 5.} Each term in the ``primed''
    six-point function~(\ref{eq:sixpt_ng}) contains a single
    delta function $\delta^3(\cdots)$.
    If the delta function argument is a small-scale wavenumber (in the
    sense defined above), then
    we assume that the six-point term under consideration
    gives a negligible contribution to $N^{(3/2)}$.
    For example, the term 
    $(Q^{gg}_{\k_1\k_3\k_6} Q^{ee}_{\k_2\k_4\k_5})$
    containing the delta function
    $\delta^3(\q+(\l-\l')/\chi_*)$ is negligible,
    whereas the term
    $(Q^{gg}_{\k_1\k_2\k_6} Q^{ee}_{\k_2\k_3\k_5})$
    containing the delta function
    $\delta^3(\k_L-\q-\q')$ is non-negligible.
    
    Rationale: In Eq.~(\ref{eq:n32_sixpoint}),
    the $N^{(3/2)}$-bias is computed by
    by integrating over small-scale wavenumbers
    $(\l/\chi_*), (\l/\chi_*)$ and large-scale
    wavenumbers $\q,\q'$.
    If the six-point function contains a term 
    such as $\delta^3(\q+(\l-\l')/\chi_*)$, this
    imposes a constraint that $(\l-\l')/\chi_*$
    be a large-scale wavenumber, which is only
    satisfied in a small part of the $(\l,\l')$-plane.
    Therefore we expect a small contribution to $N^{(3/2)}$.
\end{itemize}

Most of the six-point terms in Eq.~(\ref{eq:sixpt_factorization})
are eliminated using these criteria.
On the first line of~(\ref{eq:sixpt_factorization}),
all of the $QQ$-terms are eliminated using Approximation 5,
except $(Q^{ge}_{\k_1\k_2\k_6} Q^{ge}_{\k_3\k_4\k_5})$
which is non-negligible,
and $(Q^{ge}_{\k_1\k_2\k_5} Q^{ge}_{\k_3\k_4\k_6})$
which is a special case: it contains the delta function
$\delta^3(\k_L)$, and we neglect it since we are
interested in the $N^{(3/2)}$-bias for nonzero $k_L$.
All eight $PR$-terms on the second line of
Eq.~(\ref{eq:sixpt_factorization}) are eliminated
using Approximation 5.
Finally, the last six $S$-terms (out of eight total $S$-terms)
in Eq.~(\ref{eq:S4}) are eliminated using Approximation 4.
For example, the third term in~(\ref{eq:S4}) contains
$P_{\rm lin}(\k_1+\k_3) = P_{\rm lin}((\l-\l')/\chi_*)$,
and $(\l-\l')/\chi_*$ is small-scale (except in a
small part of the $(\l,\l')$-plane).

Summarizing this section so far, we have argued only
three terms (out of 22) in the six-point
function~(\ref{eq:sixpt_factorization}) contribute
significantly to the $N^{(3/2)}$ bias:
\begin{eqnarray}
&& \big\langle 
\delta_g(\k_1) \delta_e(\k_2) 
\delta_g(\k_3) \delta_e(\k_4)
v_r(\k_5) v_r(\k_6)
\big\rangle'_{ng} \nn \\
&& \hspace{1.5cm}
\approx 
 -\beta_1(k_2) \beta_1(k_4)
\left( \frac{k_{5r} k_{6r}}{k_5 k_6} \right)
P_{mv}(k_5) P_{mv}(k_6) 
(2\pi)^3 \delta^3(\k_1+\k_2+\k_6)
\nn \\
&& \hspace{2cm} 
+ \left( \frac{\alpha_2(k_2,k_4)}{n_h} + 
    \beta_1(k_2) \beta_1(k_4) P_{\rm lin}(\k_1+\k_2) \right)
    P_{v_r}(\k_5) (2\pi)^3 \delta^3(\k_5+\k_6)
\label{eq:sixpt_final}
\end{eqnarray}
Using this expression, we now proceed to compute the
$N^{(3/2)}$-bias, by plugging the
six-point function~(\ref{eq:sixpt_final})
into our general expression~(\ref{eq:n32_sixpoint})
for the $N^{(3/2)}$-bias, obtaining:
\begin{align}
N^{(3/2)}(\k_L)
 &= N_{v_r}^{(0)}(\k_L)^2 \frac{K_*^4}{\chi_*^8} 
 \int \frac{d^2\l}{(2\pi)^2} \frac{d^2\l'}{(2\pi)^2} 
 \frac{d^3\q}{(2\pi)^3} \, \nn \\
  & \hspace{1cm} \times \Bigg[
    \frac{P_{ge}(k_S)}{P_{gg}(k_S) C_l^{\rm tot}} 
    \frac{P_{ge}(k_S')}{P_{gg}(k_S') C_{l'}^{\rm tot}} 
    \beta_1(p) \beta_1(p') \left( \frac{q_rq'_r}{qq'} \right)
    P_{mv}(q) P_{mv}(q')
   \Bigg]_{\substack{
    \k_S = \k_L - \l/\chi_* \\ 
     \k_S' = \k_L - \l'/\chi* \\
     \q' = \k_L - \q \\
     \p = -\q + \l/\chi_* \\
     \p' = -\q' + \l'/\chi_*
   }} \nn \\
  & + N_{v_r}^{(0)}(\k_L)^2 \frac{K_*^4}{\chi_*^8} 
 \int \frac{d^2\l}{(2\pi)^2} \frac{d^2\l'}{(2\pi)^2} 
 \frac{d^3\q}{(2\pi)^3} \, \nn \\
 & \hspace{1cm} \times \Bigg[
    \frac{P_{ge}(k_S)}{P_{gg}(k_S) C_l^{\rm tot}} 
    \frac{P_{ge}(k_S')}{P_{gg}(k_S') C_{l'}^{\rm tot}} 
    \left( \frac{\alpha_2(p,p')}{n_h} + 
       \beta_1(p) \beta_1(p') P_{\rm lin}(\k_L-\q) 
     \right) P_{v_r}(\q)
   \Bigg]_{\substack{
       \k_S = \k_L - \l/\chi_* \\ 
       \k_S' = \k_L - \l'/\chi* \\
       \p = -\q + \l/\chi_* \\
       \p' = -\q + \l'/\chi_*
   }} \label{eq:n32_long1}
\end{align}
We make the following approximations inside the integrals, which
are valid for $k_L \ll k_S$:
\be
N_{v_r}(\k_L) \approx N_{v_r}(0)
\hspace{1.5cm}
k_S \approx p \approx l/\chi_*
\hspace{1.5cm}
k_S' \approx p' \approx l'/\chi_*
\ee
as in the $N^{(1)}$ case 
(see discussion near Eq.~(\ref{eq:n1_lowk_approximations})).
We also write 
$P_{\rm lin}(q') P_{v_r}(q) = P_{mv}(q') P_{mv}(q) \, q_r^2/(qq')$,
to combine the two terms in~(\ref{eq:n32_long1}) into
a single term:
\begin{align}
N^{(3/2)}(\k_L)
 &\approx N_{v_r}^{(0)}(0)^2 \frac{K_*^4}{\chi_*^8} 
 \int \frac{d^2\l}{(2\pi)^2} \frac{d^2\l'}{(2\pi)^2} 
 \frac{d^3\q}{(2\pi)^3} \, \nn \\
  & \hspace{0.5cm} \times \Bigg[
    \frac{P_{ge}(k_S)}{P_{gg}(k_S) C_l^{\rm tot}} 
    \frac{P_{ge}(k_S')}{P_{gg}(k_S') C_{l'}^{\rm tot}} 
    \left(
    \frac{\alpha_2(k_S,k_S')}{n_h} P_{v_r}(q) +
    \beta_1(k_S) \beta_1(k_S') \left( \frac{q_r^2+q_rq'_r}{qq'} \right)
    P_{mv}(q) P_{mv}(q')
    \right)
   \Bigg]_{\substack{
    k_S = l/\chi_* \\ 
     k_S' = l'/\chi* \\
     \q' = \k_L - \q \\
   }}
\end{align}
We symmetrize the integrand by replacing
$(q_r^2 + q_r q_r') \rightarrow (q_r + q_r')^2/2 = k_{Lr}^2/2$,
and use the definition of $W(l)$ in Eq.~(\ref{eq:Wl_def}),
obtaining:
\begin{align}
N^{(3/2)}(\k_L)
&\approx \int \frac{d^2\l}{(2\pi)^2} \frac{d^2\l'}{(2\pi)^2} 
 \frac{d^3\q}{(2\pi)^3} W(l) W(l') \nn \\
  & \hspace{1cm} \times \Bigg[
    \frac{1}{P_{ge}(k_S) P_{ge}(k_S')}
    \left(
    \frac{\alpha_2(k_S,k_S')}{n_h} P_{v_r}(q) +
    \beta_1(k_S) \beta_1(k_S') \left( \frac{k_{Lr}^2}{2qq'} \right)
    P_{mv}(q) P_{mv}(q')
    \right)
   \Bigg]_{\substack{
    k_S = l/\chi_* \\ 
     k_S' = l'/\chi* \\
     \q' = \k_L - \q \\
   }} \nn \\
\label{eq:n32_long2}
\end{align}
So far, our approximations should be very accurate in the limit $k_L \ll k_S$.
To simplify further, we make two more approximations that are not as precise,
but should suffice for an initial estimate of the size of $N^{(3/2)}$.
First, we assume that on kSZ scales, the galaxy-electron
power spectrum is dominated by its 1-halo term:
\be
P_{ge}(k_S) \sim P_{ge}^{1h}(k_S) = \frac{1}{\rho_m n_h} \int dM \, n(M) M u_M(k_S)
\ee
We then write some of the intermediate quantities which appear in the
integral~(\ref{eq:n32_long2}) as follows:
\begin{align}
\frac{\alpha_2(k_S,k_S')}{P_{ge}(k_S) P_{ge}(k_S')} &\sim
\frac{\big\langle M^2 u_M(k_S) u_M(k_S') \big\rangle_M}{\big\langle M u_M(k_S) \big\rangle_M \, 
\big\langle M' u_{M'}(k_S') \big\rangle_{M'}}
\label{eq:alpha_intermediate} \\
\frac{\beta_1(k_S)}{P_{ge}(k_S)} &\sim
\frac{\big\langle M b(M) u_M(k_S) \big\rangle_M}{\big\langle M u_M(k_S) \big\rangle_M}
\label{eq:beta_intermediate}
\end{align}
where we have introduced the following notation, to denote an average
over halos in the catalog:
\be
\big\langle \cdots \big\rangle_M = \frac{1}{n_h} \int_{M_{\rm min}}^\infty n(M) \, (\cdots)
\ee
Our second approximation is that the factors $u_M(k_S)$ approximately
cancel on the RHS of~(\ref{eq:alpha_intermediate}),~(\ref{eq:beta_intermediate}),
since they appear in both the numerator and denominator. Then the
right-hand sides of 
Eqs.~(\ref{eq:alpha_intermediate}),~(\ref{eq:beta_intermediate})
simplify as:
\begin{align}
\frac{\alpha_2(k_S,k_S')}{P_{ge}(k_S) P_{ge}(k_S')} \sim A
\hspace{1.5cm}
\frac{\beta_1(k_S)}{P_{ge}(k_S)} \sim B
\label{eq:alpha_beta_intermediate2}
\end{align}
where the dimensionless constants $A,B$ are defined by:
\be
A = \frac{\langle M^2 \rangle_M}{\langle M \rangle_M^2}
\hspace{1.5cm}
B = \frac{\langle M b(M) \rangle_M}{\langle M \rangle_M}
\label{eq:AB_def}
\ee
Making the approximations~(\ref{eq:alpha_beta_intermediate2})
in Eq.~(\ref{eq:n32_long2}), the $N^{(3/2)}$-bias simplifies
significantly:
\begin{align}
N^{(3/2)}(\k_L)
 & \sim
 \int \frac{d^2\l}{(2\pi)^2} \frac{d^2\l'}{(2\pi)^2} 
 \frac{d^3\q}{(2\pi)^3} \, W(l) W(l')
 \left[ 
   \frac{A}{n_h} P_{v_r}(q) + 
   \frac{B^2 k_{Lr}^2}{2qq'} P_{mv}(q) P_{mv}(q') 
  \right]_{\q'=\k_L-\q} \nn \\
 &= \frac{A}{n_h} \langle v_r^2 \rangle
     + \frac{B^2 k_{Lr}^2}{2}
      \int \frac{d^3\q}{(2\pi)^3} \,
      \left[ 
          \frac{P_{mv}(q) P_{mv}(q')}{qq'}
      \right]_{\q'=\k_L-\q}  \label{eq:n32_penultimate}
\end{align}
where in the second line, we have used $\int d^2\l/(2\pi)^2 \, W(l) = 1$,
and $\langle v_r^2 \rangle$ denotes the variance of the radial velocity field:
\be
\langle v_r^2 \rangle \equiv
\int \frac{d^3\k}{(2\pi)^3} P_{v_r}(k) =
\int \frac{k^2 \, dk}{6\pi^2} P_v(k)
\ee
Finally, we note that in the 3-d integral~(\ref{eq:n32_penultimate}),
one angular integral can be done analytically, reducing to a 2-d
integral. We omit the details and quote the final result:
\be
N^{(3/2)}(\k_L)
 \sim \frac{A}{n_h} \langle v_r^2 \rangle
    + \frac{B^2 k_{Lr}^2}{8\pi^2 k_L} \int_0^\infty dq
      \int_{|k_L-q|}^{k_L+q} dq' \,
      P_{mv}(q) P_{mv}(q')  \label{eq:n32_final}
\ee
To angle-average over $\k_L$ (as we will do in
Figure~\ref{fig:N3/2 bias} shortly), we replace
$k_{Lr}^2 \rightarrow k_L^2/3$ in the second term.
The $A$-term in Eq.~(\ref{eq:n32_final}) is constant in $k_L$,
and the $B$-term goes to zero at both low and high $k_L$, with
a peak at $k_L \sim 0.03$ Mpc$^{-1}$.

\subsection{Numerical evaluation and discussion}
\label{ssec:bias_numerics}

In the last few sections, we identified several new contributions
to the kSZ reconstruction noise, going beyond the $N^{(0)}$-bias
from~\cite{Smith:2016lnt}.
Can these new contributions explain the excess noise in our
simulations, shown previously in Figure~\ref{fig:quijote_mean_powerspectrum}?

In Figure~\ref{fig:N3/2 bias}, we numerically evaluate the
$N^{(0)}$, $N^{(1)}$, and $N^{(3/2)}$ biases as follows.
All power spectra are angle-averaged over $\k$.
We compute the $N^{(0)}$-bias using
Eqs.~(\ref{eq:nvr_fast}),~(\ref{eq:f1f2}),
but to maximize consistency with our simulations,
we replace integrals (either $\int d^3\x$,
$\int d^3\k$, or $\int d^2\l$) by sums over the
discrete set of pixels (or Fourier modes) 
used in our simulation pipeline.

We compute the angle-averaged $N^{(1)}$-bias using Eq.~(\ref{eq:n1_avg}),
and the $N^{(3/2)}$-bias using Eq.~(\ref{eq:n32_final}).
Note that~(\ref{eq:n1_avg}) and~(\ref{eq:n32_final}) are
approximations which are accurate for $k \rightarrow 0$.
To evaluate~(\ref{eq:n32_final}), we need numerical values for the
constants $A,B$ defined in Eq.~(\ref{eq:AB_def}).
We get $A=2.3$ using the measured halo mass function from our $N$-body 
simulations. We approximate $B \sim b_g$, where
$b_g = 3.24$ is the halo bias of our simulations. (This is an
approximation since $b_g$ is calculated weighting all
halos equally, whereas $B$ is the mass-weighted halo bias.)

Our first result in Figure~\ref{fig:N3/2 bias} is that the
$N^{(1)}$-bias is negligible.
As a check on our $N^{(1)}$ calculation, we compared to
Gaussian Monte Carlo simulations which are designed to
isolate the $N^{(1)}$-bias, and find good agreement.
In more detail,
each Gaussian simulation consists of 3-d Gaussian
fields $v_r, \delta_g, \delta_e$ with the same auto and
cross power spectra as the Quijote simulations.
For each triple $(i,j,k)$ of Gaussian simulations, let
$\hv_r^{ijk}$ denote the kSZ velocity reconstruction
using fields $v_r^i$, $\delta_g^j$, $\delta_e^k$ from
simulations $i,j,k$.
Then the cross power spectrum between $\hv_r^{ijk}$
and $\hv_r^{ikj}$ is equal to $N^{(1)}$, with no
$N^{(0)}$ or $N^{(3/2)}$ contribution, since $N^{(1)}$ 
is the only surviving contraction in 
Eq.~(\ref{eq:pvr_wick_contractions}).

Our main result in Figure~\ref{fig:N3/2 bias} is that the
$N^{(3/2)}$-bias agrees well with the excess noise seen 
in simulations!
(Surprisingly, the agreement holds to high $k$, even
though we have freely made approximations which are
only valid for $k\rightarrow 0$.)
Our conclusion is that higher-order biases are real,
non-negligible contributions to kSZ reconstruction noise
which can be calculated systematically in the halo model.

\begin{figure}
    \includegraphics[]{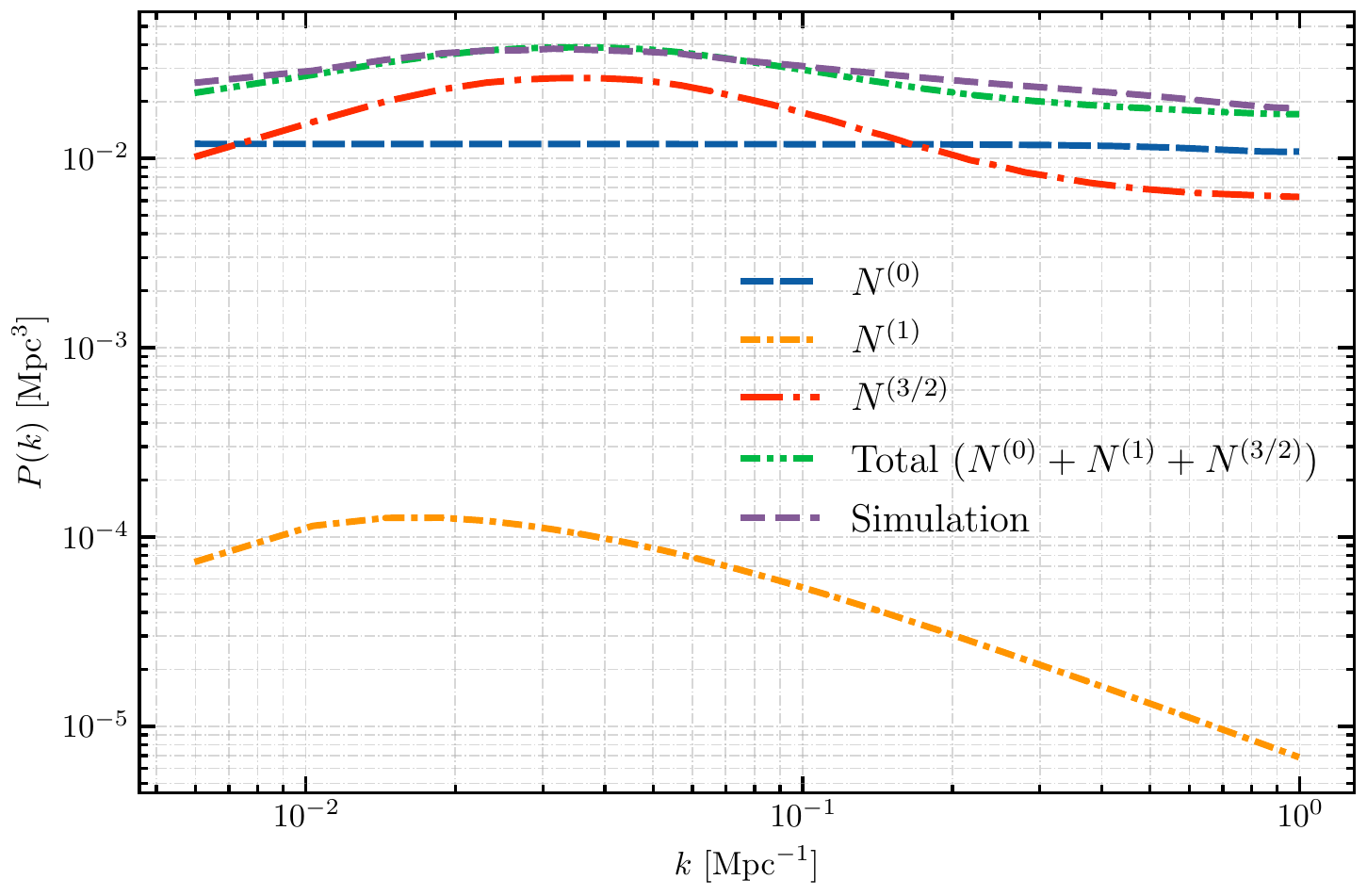}
    \caption{Contributions to the KSZ reconstruction noise, computed
    as described in~\S\ref{ssec:bias_numerics}.
    The reconstruction noise in simulations agrees well with the sum
    of analytic contributions $(N^{(0)} + N^{(1)} + N^{(3/2)})$.
    All noise power spectra have been angle-averaged over $\k$,
    and $N^{(1)}$ and $N^{(3/2)}$ have been computed using approximations
    which are valid for $k \rightarrow 0$.}
    \label{fig:N3/2 bias}
\end{figure}

The preceding results have assumed fiducial survey
parameters from~\S\ref{sec:pipeline}.
In this paper, we will not explore dependence on
galaxy density $n_g$ or redshift $z$,
although this should be fairly straightforward using
our expressions for $N^{(0)}$ and $N^{(3/2)}$ bias.
However, one parameter which is easy to analyze is the
CMB noise level.
In the approximation~(\ref{eq:n32_final}), the $N^{(3/2)}$-bias
is independent of the CMB noise. On the other hand,
Eq.~(\ref{eq:nvr_fourier}) shows that the $N^{(0)}$-bias
is proportional to $C_l^{\rm tot} = (C_l + N_l)$ evaluated on 
kSZ scales $l\sim 5000$.
Therefore, as the CMB experiment becomes more sensitive,
the $N^{(3/2)}$-bias becomes more important, relative to $N^{(0)}$.

Since our simulations use futuristic CMB noise parameters
(0.5 $\mu$K-arcmin, $\theta_{\rm fwhm}$ = 1 arcmin), and 
galaxy survey parameters comparable to DESI, it seems likely 
that $N^{(3/2)}$ will be small (relative to $N^{(0)}$)
for DESI in combination with near-future CMB experiments such as 
Simons Observatory.
However, if DESI is replaced by an experiment with larger
galaxy density (e.g.~Rubin Observatory), or if the CMB
noise is $\lesssim 1$ $\mu$K-arcmin, then $N^{(3/2)}$
may be important.

\section{Recovering $f_{NL}$ with an MCMC pipeline}
\label{sec:fnl_results}

In this section, we develop an MCMC-based analysis pipeline which recovers
the value of $f_{NL}$ from a galaxy catalog and CMB map.
We demonstrate the ability of our pipeline to recover the correct value of $f_{NL}$,
and validate its statistical errors with Monte Carlo simulations.

In our pipeline, $f_{NL}$ sensitivity arises entirely from
$f_{NL}$ dependence of the galaxy bias: $b(k) = b_g + f_{NL} b_{ng} / k^2$.
The velocity reconstruction $\hv_r$ is not directly $f_{NL}$-sensitive.
However, $\hv_r$ can be used to cancel sample variance in 
the galaxy field, thus improving
the statistical error on $f_{NL}$ relative to a measurement of $\delta_g$ alone.
The idea of sample variance cancellation was introduced by Seljak in~\cite{Seljak_2009}.
Sample variance cancellation is automatically incorporated by our MCMC pipeline,
since we write down the full posterior likelihood $\L(f_{NL} | \delta_g, \hv_r)$
(Eq.~(\ref{eq:likelihood_ksz}) below), which includes sample variance cancellation
automatically.

When constructing our posterior likelihood, we assume 
that the reconstruction noise power spectrum is 
given by the $N^{(0)}$-bias in Eq.~(\ref{eq:nvr_fourier}).
This neglects the $N^{(3/2)}$ bias, even though we
have shown that $N^{(3/2)}$ is comparable to $N^{(0)}$
for our fiducial survey parameters.
In principle, neglecting $N^{(3/2)}$ can produce both biased
$f_{NL}$ estimates and underestimated statistical errors
(as in the CMB lensing case).
However, in this section we will find that within
statistical errors of our simulations, our MCMC pipeline
recovers unbiased estimates of $f_{NL}$, with scatter
consistent with a Fisher matrix forecast.

In our pipeline, we have perfect knowledge of the galaxy-electron
power spectrum $P_{ge}(k_S)$, and therefore we expect the reconstruction
bias $b_v$ to equal 1.
However, in our MCMC's, we will include $b_v$ as a nuisance parameter
and marginalize it, so that our analysis is more representative of real 
experiments.
As a consistency check, we expect the value of $b_v$ 
recovered from the MCMC's to be consistent with 1.

\subsection{MCMC pipeline description}

The inputs to our pipeline are a realization $\delta_h(\k)$
of the 3-d halo field, and the kSZ velocity reconstruction $\hv_r(\k)$.
We want to constrain the cosmological parameter $f_{NL}$, and the nuisance
parameters $b_g, b_v$.
Here, $b_g$ is the Gaussian halo bias, and $b_v$ is the kSZ velocity reconstruction bias
from~\S\ref{ssec:vrec_results}.
For notational compactness, let $\pi$ denote the three-component parameter
vector $\pi = (f_{NL}, b_g, b_v)$.

We start by writing down the two-point statistics of the fields $\delta_g$ and $\hv_r$.
For each Fourier mode $\k$, let $\theta(\k)$ be the two-component vector of fields:
\be
\theta(\k) = \left( \begin{array}{c}
  \delta_h(\k) \\
  \hv_r(\k)
\end{array} \right)
\ee
Let $C(\k,\pi)$ be the 2-by-2 Hermitian matrix defined by:
\be
\big\langle \theta(\k) \, \theta(\k')^\dag \big\rangle = C(\k,\pi) \, (2\pi)^3 \delta^3(\k-\k')
\ee
We model $C(\k,\pi)$ on large scales by:
\ba
C_{11}(k,\pi) &=& b_h(k,\pi)^2 P_{\rm lin}(k) + \frac{1}{n_h} \label{eq:c11} \\
C_{12}(\k,\pi) &=& -ik_r \left( \frac{faH}{k^2} \right) \, b_v \, b_h(k,\pi) P_{\rm lin}(k) \label{eq:c12} \\
C_{22}(\k,\pi) &=& k_r^2 \left( \frac{faH}{k^2} \right)^2 \, b_v^2 \, P_{\rm lin}(k) + N_{v_r}^{(0)}(\k) \label{eq:c22}
\ea
where $b_h(k,\pi)$ is the non-Gaussian halo bias:
\be
b_h(k,\pi) = b_g + f_{NL} \frac{2 \delta_c (b_g-1)}{\alpha(k,z)}
\hspace{1cm} (\delta_c=1.42)
\ee
and $N^{(0)}_{v_r}(\k)$ was given in Eq.~(\ref{eq:nvr_fourier}).
The model for $C(\k,\pi)$ in Eqs.~(\ref{eq:c11})--(\ref{eq:c22})
follows if we assume that $\delta_h$ and $\hv_r$ are modeled as:
\ba
\delta_h(\k) &=& b_h(k) \, \delta_m(\k) + \mbox{(Poisson noise)} \nn \\
\hv_r(\k) &=& i b_v k_r \frac{faH}{k} \delta_m(\k)
  + \mbox{(Reconstruction noise)}
\ea
In the previous section, we tested these assumptions systematically,
thus validating our model~(\ref{eq:c11})--(\ref{eq:c22}) for the 
two-point function $C(\k,\pi)$.

However, to run an MCMC we need to go beyond the two-point function, by
writing down a model for the posterior likelihood $\L(\pi | \theta)$
for parameter vector $\pi$, given data realization $\theta(\k)$.
Here, we simply assume the Gaussian likelihood derived from
the two-point function in Eqs.~(\ref{eq:c11})--(\ref{eq:c22}):
\be
\L\big(\pi | \theta) \propto \prod_\k
  \big( \mbox{Det} \, C(\k,\pi) \big)^{-1/2}
  \exp\left( -\frac{\theta(\k)^\dag C(\k,\pi)^{-1} \theta(\k)}{2V} \right)  \label{eq:likelihood_ksz}
\ee
where the survey volume $V$ on the RHS arises from our 
finite-volume Fourier convention in Eq.~(\ref{eq:fourier_3d_finite_vol}).
This ``field-level'' likelihood function makes fewer approximations
than a likelihood function based on power spectrum bandpowers.
However, we emphasize that the likelihood~(\ref{eq:likelihood_ksz})
treats $\delta_g$ and $\hv_r$ as Gaussian fields, and results from 
previous sections do not imply its validity.
Indeed, the main purpose of this section is to validate 
the Gaussian likelihood function, by showing that it leads to
valid constraints on $f_{NL}$.

We truncate the likelihood~(\ref{eq:likelihood_ksz})
at $k_{\rm max} = 0.012$ Mpc$^{-1}$.
The posterior likelihood is sampled using Goodman-Weare sampling 
algorithm~\cite{goodman2010} implemented in the public library 
\texttt{emcee}~\cite{Foreman_Mackey_2013}.
We use flat priors over a reasonable range of values for all
three model parameters of the model, and run the chain long enough
to fulfil recommended convergence criterion based on correlation length.

\subsection{Unbiased $f_{NL}$ estimates from MCMC}

We now present results from running our MCMC pipeline on $N$-body simulations.
First, we check for additive bias in $f_{NL}$, by confirming that when
the MCMC pipeline is run on simulations with $f_{NL}=0$, there is no
bias toward positive or negative $f_{NL}$.

\begin{figure}
    \centerline{\includegraphics[]{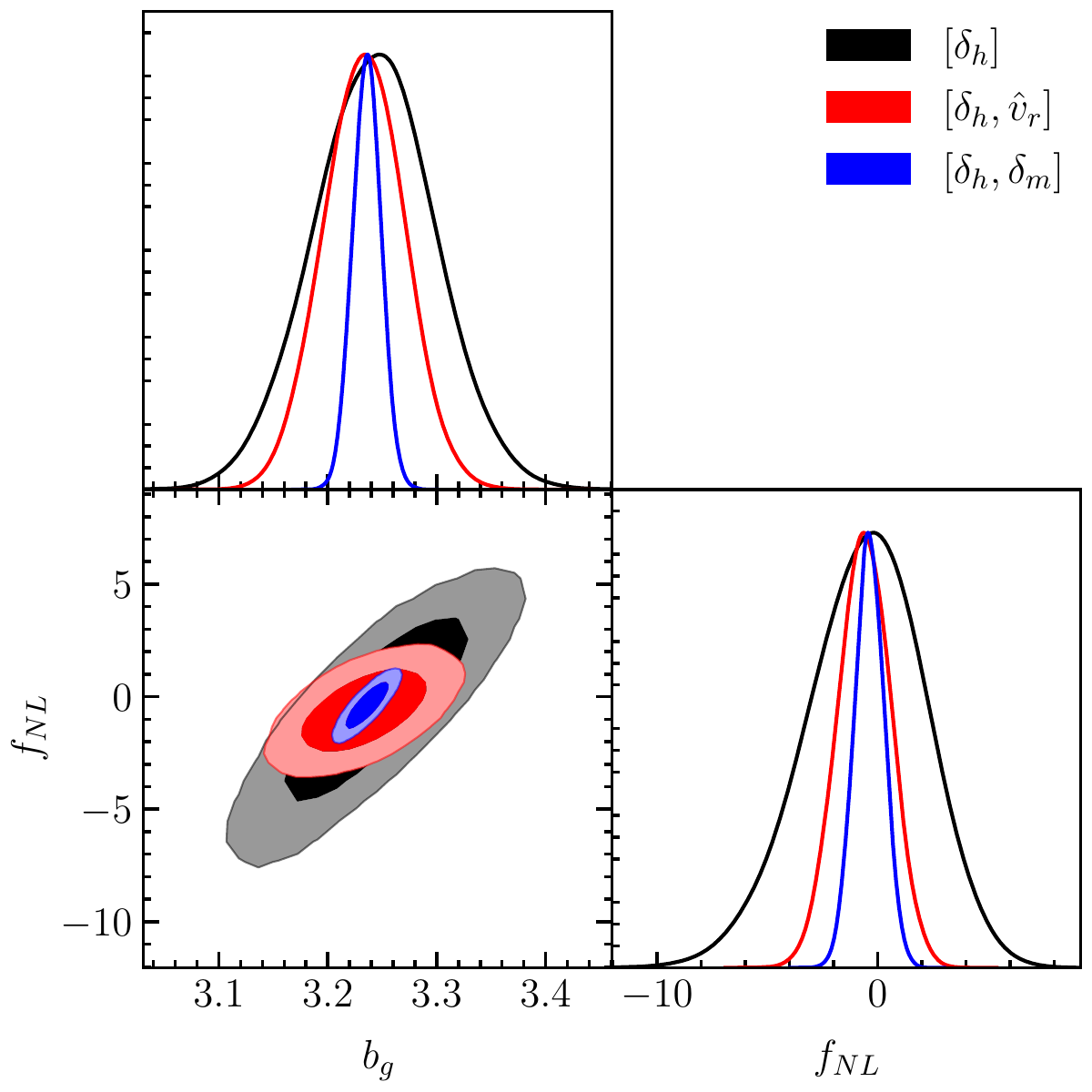}}
     \caption{MCMC posteriors on $(b_g, f_{nl})$ from combined analysis of 100
     high resolution Quijote simulations with $f_{NL}=0$. 
     The three likelihoods correspond to MCMC analysis of the halo field alone ($\delta_h$),
     joint analysis of the halo field and kSZ velocity reconstruction ($\delta_h, \hv_r$),
     and joint analysis of the halo field and the noise-free matter field ($\delta_h, \delta_m$).
     In the second case $(\delta_h, \hv_r)$, likelihoods have been marginalized over the
     additional nuisance parameter $b_v$.}
    \label{fig:mcmc_combined}
\end{figure}

In Figure~\ref{fig:mcmc_combined}, we jointly analyze all 100
Quijote simulations with $f_{NL}=0$, by multiplying together their
posterior likelihoods.
We run three versions of the MCMC pipeline as follows.
First, we constrain parameters using the halo field alone $(\delta_h)$.
Second, we use our standard setup described in the previous section,
where we include the halo field and the kSZ velocity reconstruction 
$(\delta_h + \hv_r)$.
Third, we use the halo field and a perfect, noise-free realization of
the matter overdensity $(\delta_h + \delta_m)$.
Note that in the second case $(\delta_h + \hv_r)$, the MCMC parameters
are $(f_{NL}, b_g, b_v)$, whereas in the first and third cases, the
parameters are $(f_{NL}, b_g)$.
In the second case $(\delta_h + \hv_r)$, the likelihoods in 
Figure~\ref{fig:mcmc_combined} are marginalized over the additional
parameter $b_v$.

The $f_{NL}$ constraint in Figure~\ref{fig:mcmc_combined} 
from $(\delta_h + \hv_r)$ is significantly better than 
the $\delta_h$-only constraint, and slightly
worse than the $(\delta_h + \delta_m)$-constraint.
This shows that sample variance cancellation between
$\delta_h$ and $\hv_r$ is happening, and the level of
cancellation is comparable to what
would be obtained from a perfect measurement of $\delta_m$.

From Figure~\ref{fig:mcmc_combined}, we can also conclude that
the $f_{NL}$ estimates from our MCMC pipeline are not additively biased.
The combined $(\delta_h + \hv_r)$ likelihood is consistent 
with $f_{NL}=0$, within the statistical error from
100 simulations.
Any additive $f_{NL}$ bias must be smaller than this statistical
error (roughly $\Delta f_{NL}=2$).

\begin{figure}
    \centerline{\includegraphics[scale=0.9]{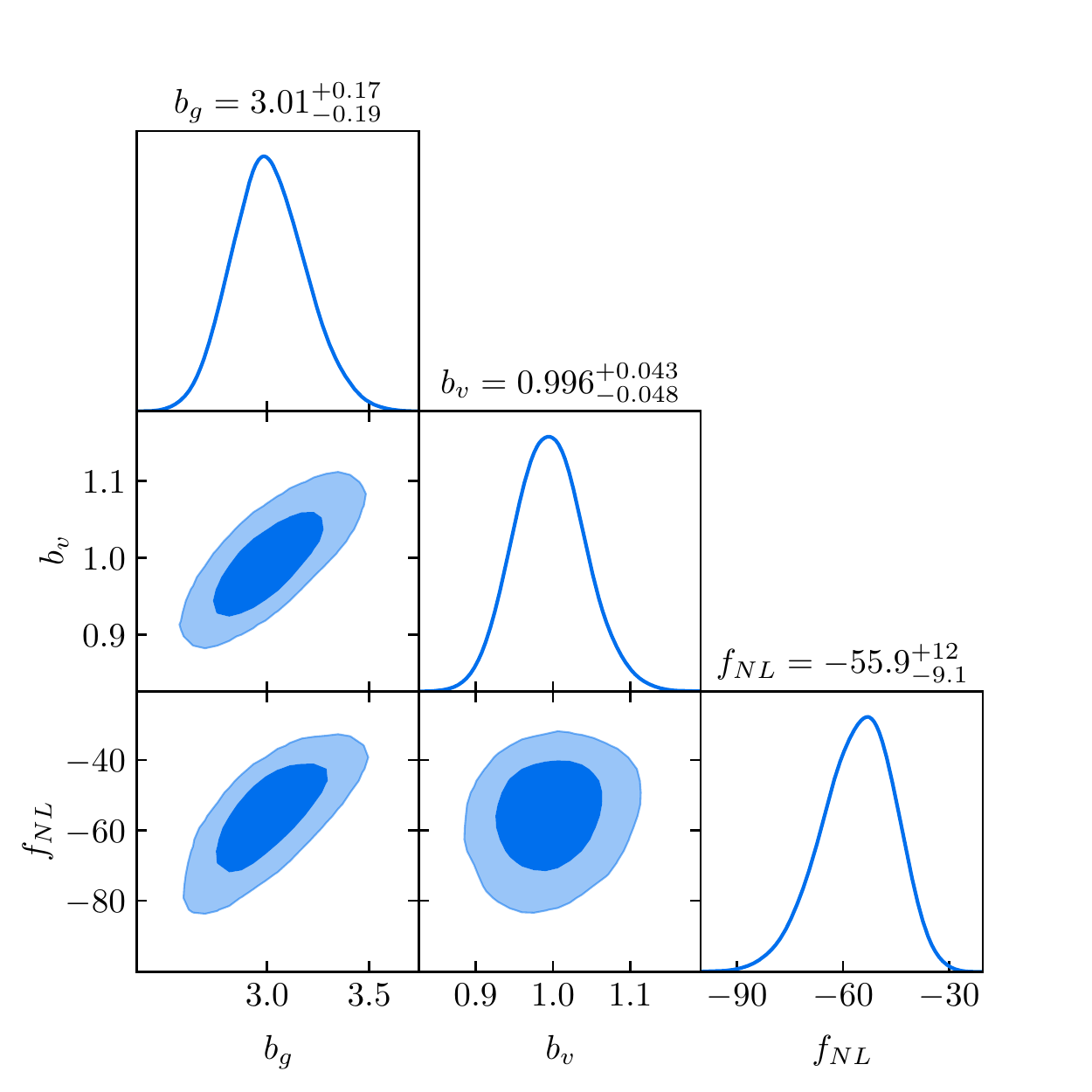}}
    \vspace{0.5cm}
    \centerline{\includegraphics[scale=0.9]{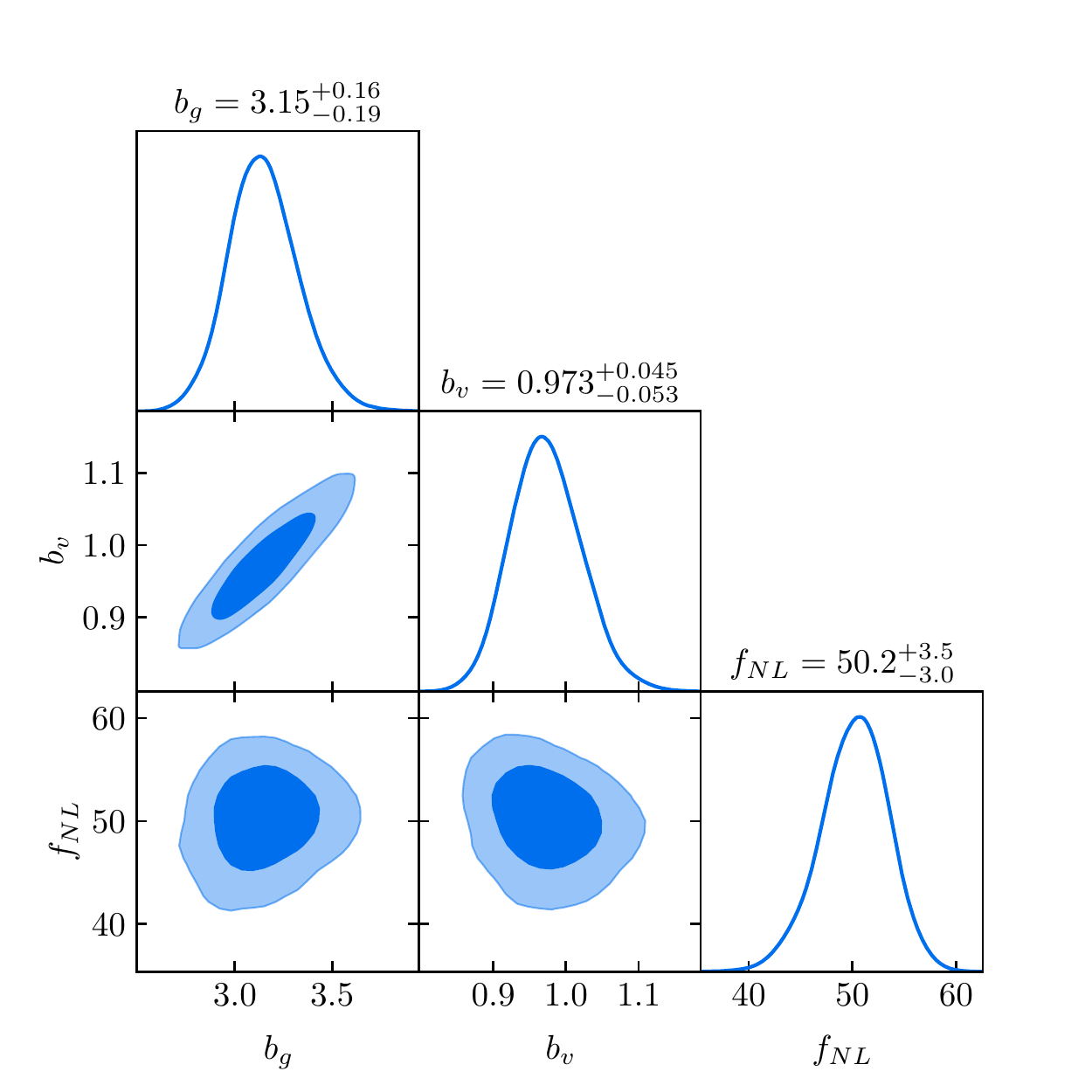}}
    \caption{MCMC constraints for $f_{nl}=-50$ (top panel) and $f_{NL}=50$ (bottom).
    Each panel combines likelihoods from four $N$-body simulations, each with volume
    1 $h^{-3}$ Gpc$^3$. The recovered $f_{NL}$ values are consistent with the true
    values, within statistical errors.}
    \label{fig:mcmc_fnl}
\end{figure}

Next, we check for multiplicative bias in $f_{NL}$, by analyzing
simulations with $f_{NL} \ne 0$ and confirming that we recover the
correct value of $f_{NL}$.
In Figure~\ref{fig:mcmc_fnl}, we present results from non-Gaussian $N$-body
simulations with $f_{NL} = \pm 50$.
It is seen that the MCMC pipeline recovers the correct value of $f_{NL}$ within
its reported statistical error (around 10--20\%).
The total simulation volume is smaller (8 $h^{-3}$ Gpc$^3$) here than in the $f_{NL} = 0$
case (100 $h^{-3}$ Gpc$^3$), where Quijote simulations are available.
Therefore, we cannot characterize the behavior of the MCMC pipeline as precisely
as we can in the $f_{NL}=0$ case.
However, the current observational situation is that $f_{NL}$ has not 
been detected, and the priority for upcoming experiments will be testing the
null hypothesis that $f_{NL}=0$.
In this situation, it should suffice to have a precise characterization
of the pipeline on simulations with $f_{NL}=0$, and a $\approx$10-20\%
test for bias on simulations with nonzero $f_{NL}$.

\subsection{Consistency between MCMC results and Fisher matrix forecasts}

\begin{figure}
    \centerline{\includegraphics[]{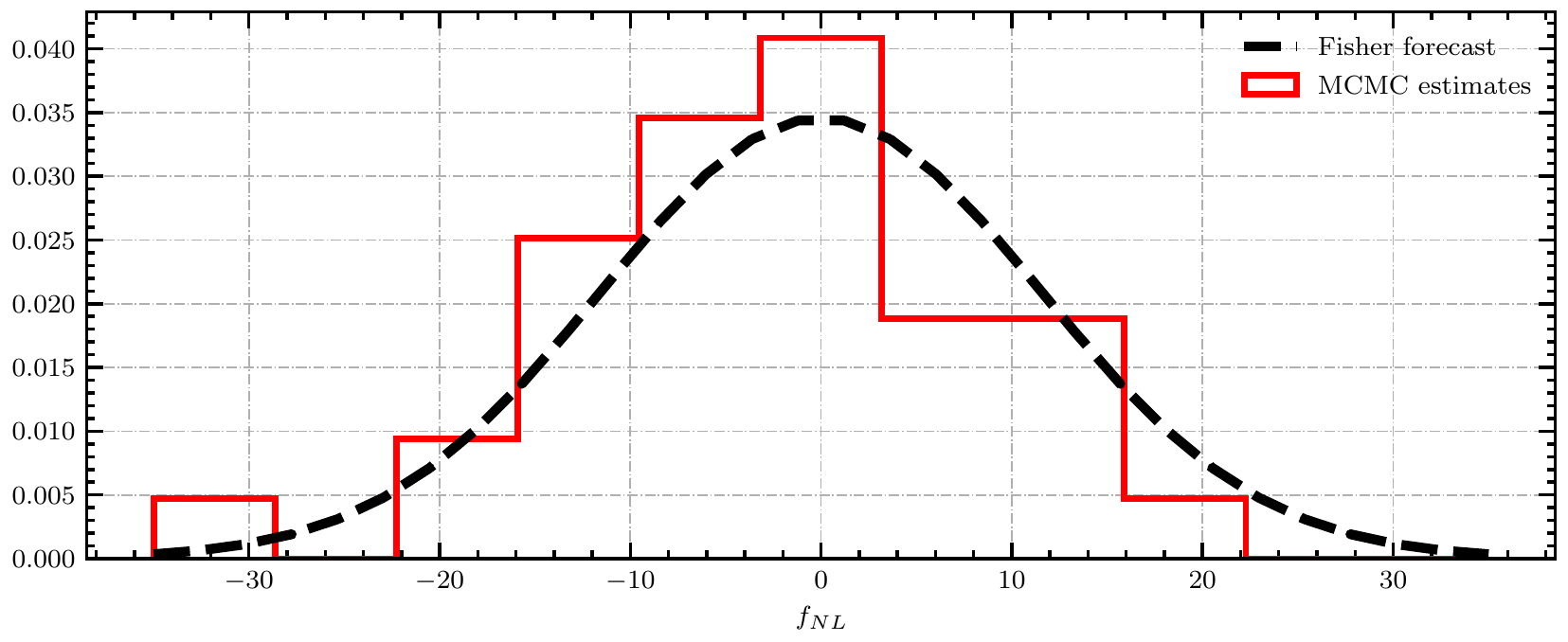}}
    \caption{A test of the error estimates from our MCMC pipeline.
    The solid histogram contains one $f_{NL}$ estimate from each of 100 high-res
    Quijote simulations with $f_{NL}=0$, obtained by taking the median of the
    $f_{NL}$ posterior likelihood (after marginalizing over $b_g,b_v$).
    The dashed line is a Gaussian whose width is equal to the Fisher
    forecasted error on $f_{NL}$.
    The two distributions have equal widths, within statistical
    errors from 100 simulations.}
    \label{fig:mcmc_medians}
\end{figure}

The tests in the previous section show that the MCMC pipeline
recovers unbiased estimates of $f_{NL}$, but do not test statistical errors on $f_{NL}$ inferred from the posteriors.
In this section, we will validate $f_{NL}$ errors from the MCMC pipeline.

For the sake of discussion, we briefly describe a completely rigorous,
Bayesian procedure for validating $f_{NL}$ errors (even though this is not
what we will end up doing!)
Suppose we choose a prior $p(f_{NL})$, and generate a large number of simulations
with $f_{NL}$ values sampled from the prior.
For each simulation $s$, we use MCMC to compute the posterior likelihood
$p(f_{NL}|s)$, and rank the true value of $f_{NL}$ within the posterior
likelihood, to obtain a quantile $0 < q < 1$.
Then we should find that $q$ is uniform distributed, if the posterior likelihoods
have been computed correctly.
This is a precise statement that can be proved rigorously.
This check validates error estimates from the MCMC pipeline, in the
sense that if the MCMC pipeline overestimates its error bars (i.e.~returns
posterior likelihoods which are too wide), then the distribution of $q$-values
will be narrower than uniform.

The difficulty with this method is that it would require many simulations
with $f_{NL} \ne 0$, which would be very expensive.
Instead, we will use an alternative method which uses only simulations
with $f_{NL}=0$ (so that we can use the Quijote simulations).
For each such simulation, let $\fnlmed$ be the
median of the MCMC posterior likelihood for $f_{NL}$ (marginalized
over $b_g,b_v$).
Let $\sigma(\fnlmed)$ be the RMS scatter in $\fnlmed$ over
100 Quijote simulations.
We will compare $\sigma(\fnlmed)$ to the Fisher forecasted statistical
error on $f_{NL}$ (which we will denote $\sigma_F(f_{NL})$).
Intuitively, we expect that 
$\sigma(\fnlmed) \approx \sigma_F(f_{NL})$,
but this is not rigorously guaranteed, so this test is not 
quite as precise as the Bayesian test described above.
However, the Cram\'er-Rao inequality implies
$\sigma(\fnlmed) \ge \sigma_F(f_{NL})$.

We briefly describe the implementation of our Fisher
matrix forecast. The 3-by-3 Fisher matrix is given by:
\be
F_{ab} = \frac{1}{2} \sum_{\k} \mbox{Tr} \left[ 
  C(\k,\pi)^{-1} \frac{\partial C(\k,\pi)}{\partial \pi_a}
  C(\k,\pi)^{-1} \frac{\partial C(\k,\pi)}{\partial \pi_b} \right]
\ee
where $a,b$ index elements of the parameter vector
$(\pi_1,\pi_2,\pi_3) = (f_{NL}, b_g, b_v)$.
The 2-by-2 covariance matrix $C(\k,\pi)$ was defined
previously in Eqs.~(\ref{eq:c11})--(\ref{eq:c22}), and
parameter derivatives of $C$ are straightforward to compute.
The Fisher-forecasted statistical error on $f_{NL}$,
marginalized over $(b_g,b_v)$, is given by
$\sigma_F(f_{NL}) = \sqrt{(F^{-1})_{11}}$.

In Figure~\ref{fig:mcmc_medians}, the solid histogram shows
values of $\fnlmed$ for all 100 Quijote simulations.
The dashed curve is a Gaussian whose width is equal to the Fisher
forecasted error on $f_{NL}$.
We find that $\sigma(\fnlmed) = 10.54$ and $\sigma_F(f_{NL}) = 11.54$.
These values are equal (at $2\sigma$) within statistical errors
from 100 Quijote simulations.
This agreement was not guaranteed in advance, since the Fisher
forecast makes approximations (neglecting $N^{(3/2)}$-bias,
treating $\delta_g$ and $\hv_r$ as Gaussian fields), whereas
$\sigma(\fnlmed)$ is a Monte Carlo error estimate based on
$N$-body simulations.
The observed agreement directly valdiates previous Fisher 
forecasts based on kSZ velocity reconstruction 
(e.g.~\cite{Smith:2018bpn,Munchmeyer:2018eey}).

\section{Discussion}
\label{sec:discussion}

KSZ velocity reconstruction is a promising method for constraining cosmology.
However, almost all work to date (with the notable exception of~\cite{Cayuso_2018})
has been based on analytic modeling which has not been tested with
simulations.
In this paper, we have made a detailed comparison between analytic
models and $N$-body simulations.
Overall, we have found good agreement, concluding with an end-to-end pipeline
which recovers unbiased estimates of $f_{NL}$ from simulated galaxy and kSZ
datasets, with statistical errors which are consistent with a Fisher matrix
forecast. This initial study is a starting point for future refinements, and
we list some possibilities here:

\begin{itemize}

\item 
We have found a discrepancy between velocity reconstruction
noise in our $N$-body simulations, and the kSZ $N^{(0)}$-bias which is
typically used in forecasts.
Using the halo model, we revisited the calculation
of the reconstruction noise power spectrum, and found new terms: the kSZ
$N^{(1)}$ and $N^{(3/2)}$ biases.
We computed these terms numerically and found that $N^{(1)}$ is negligible,
while $N^{(3/2)}$ matches the excess noise seen in simulations 
(Figure~\ref{fig:N3/2 bias}).
Our final expression for $N^{(3/2)}$ (Eq.~(\ref{eq:n32_final})) is
algebraically simple enough that it should be straightforward
to include in future forecasts or data analysis.

\item 
Similarly, we have found that the non-Gaussian bandpower covariance
of the reconstruction noise can be large (\S\ref{ssec:bandpower_covariance}).
It would be interesting to model this effect, e.g. using the halo model.

For our choice of fiducial survey parameters (\S\ref{sec:pipeline}), 
neither the non-Gaussian
bandpower covariance nor the $N^{(3/2)}$-bias 
has much impact on the bottom-line $f_{NL}$ constraint.
However, this may not be the case for other
choices of survey parameters (CMB noise, galaxy density,
redshift, etc.), and systematic exploration of parameter
dependence would be valuable.

\item
We have used collisionless $N$-body simulations, making the approximations
that electrons trace dark matter ($\delta_e = \delta_m$) and galaxies
are in one-to-one correspondence with dark matter halos ($\delta_g = \delta_h$).
These are crude approximations, and our simulations
overpredict the small-scale galaxy-electron power
spectrum $P_{ge}(k_S)$ by an order-one factor.
We do not think this is an issue for purposes of this paper, where
our goal is to test agreement between simulations and theory under
self-consistent assumptions.
However, it would be good to check this by incorporating baryonic physics,
for example using the Illustris-TNG simulation~\cite{Nelson:2018uso}.

\item
We have used a snapshot geometry (\S\ref{ssec:pipeline_geometry}), which
could be generalized to a lightcone geometry with redshift evolution.

\item
We have not included CMB foregrounds and other non-Gaussian secondaries
(e.g.~lensing). This issue is not as serious as it sounds, since there
are symmetry arguments which show that the velocity reconstruction bias
produced by foregrounds and secondaries should be small.

In the case of CMB lensing, there is a symmetry which reverses the
sign of the primary CMB anisotropy $T_{\rm pri} \rightarrow -T_{\rm pri}$
while leaving late-universe LSS unchanged.
Strictly speaking, this is an approximate symmetry which assumes
that the last scattering surface and the late universe are
statistically independent, but this is an excellent approximation
on small scales.
Under this symmetry, the lensed CMB is odd ($T_{\rm len} \rightarrow -T_{\rm len})$,
whereas the kSZ and other secondaries/foregrounds are even ($T \rightarrow T$).
This implies that lensing cannot produce a velocity reconstruction bias 
$\langle \hv_r \rangle$.

Most non-kSZ secondaries (including CMB lensing, but also e.g.~tSZ
or CIB) are even under radial reflection symmetry, whereas the kSZ is odd.
This implies that there is no velocity reconstruction bias.
However, radial reflection is only an approximate
symmetry in a lightcone geometry (unlike the snapshot geometry 
where it is exact), so there
will be some residual bias which should be quantified with simulations.
Additionally, even if foregrounds/secondaries produce minimal velocity
reconstruction bias, their non-Gaussian statistics may produce extra
reconstruction noise (relative to a Gaussian field), and it would be
useful to quantify this with simulations.

\item
A natural extension of this work would be to study the effect
of redshift space distortions (RSD's) or photometric redshift errors.
Ref.~\cite{Smith:2018bpn} makes analytic predictions
for the effect of RSD's and photo-$z$ errors on kSZ velocity
reconstruction, on large scales and assuming a simplified photo-$z$
model.
It would be interesting to compare these predictions to simulations.
Additionally, simulations could be used to study small-scale RSD's
(``Fingers of God'') and catastrophic photo-$z$ errors,
where analytic predictions are difficult.

If RSD's are included in the simulations, then it should be possible
to break the kSZ optical depth degeneracy, as first proposed 
in~\cite{Sugiyama:2016rue}. More precisely, the $\langle g v_r \rangle$ 
correlation function contains terms proportional to $\mu^0$ and $\mu^2$,
and by comparing the amplitude of these terms, the parameter combination
$f/b_g$ can be constrained, with no contribution from $b_v$.
It would be very interesting to test this picture with simulations.

\item
We have focused on constraining $f_{NL}$, and it would
be interesting to study other applications of kSZ tomography, for example
using sample variance cancellation to constrain the RSD parameter 
$f = \partial \log D / \partial \log a$. Similarly, we could generalize
the non-Gaussian model, by introducing scale-dependent $f_{NL}$, or
the ``$g_{NL}$ model'' with $\zeta^3$-type non-Gaussianity.
\end{itemize}

\section*{Acknowledgements}

We thank Niayesh Afshordi, Neal Dalal, Simone Ferraro, Matt Johnson, Mathew Madhavacheril,
Moritz M\"unchmeyer, and Emmanuel Schaan for discussions.
KMS was supported by an NSERC Discovery Grant, an Ontario Early Researcher Award, a CIFAR fellowship,
and by the Centre for the Universe at Perimeter Institute.
Research at Perimeter Institute is supported by the Government of Canada
through Industry Canada and by the Province of Ontario through the Ministry of Research \& Innovation.

\bibliography{ksz_nbody}

\appendix

\section{Diagrammatic rules for the halo model}
\label{app:diagrams}

The main purpose of this appendix is to derive
Eqs.~(\ref{eq:sixpt_factorization})--(\ref{eq:S4})
for the non-Gaussian six-point function
\be
\big\langle \delta_g(\k_1) \delta_e(\k_2) \delta_g(\k_3)
\delta_e(\k_4) v_r(\k_5) v_r(\k_6)
\big\rangle_{ng} \label{eq:diag6}
\ee
in the halo model.
In general, $n$-point correlation functions in the halo model consist
of many combinatorial terms.
A second purpose of this appendix is to show that these terms can be
enumerated using diagrammatic rules, similar to Feynman rules in QFT.

We consider the simplest version of the halo model, in which halos
are linearly biased tracers of a Gaussian field $\delta_{\rm lin}$
(the linear density field).
In this model, the expected number of halos per volume per unit
halo mass is:
\be
s(M,\x) = n(M) \big( 1 + b(M) \delta_{\rm lin}(\x) \big)  \label{eq:s_def}
\ee
where $n(M)$ is the halo mass function and $b(M)$ is the linear bias.
We will call $s(M,\x)$ the {\em halo source field}, and distingiush
it from the halo density field $\delta_h(\x)$, which is a sum of delta
functions.
By assumption in the halo model, the halo density field is given
by Poisson-sampling the halo source field.

We consider fields $\delta_X$ which are sums over halos:
\be
\delta_X(\k) = \sum_j W_X(M_j,k) e^{-i\k\cdot\x_j}  \label{eq:delta_X}
\ee
where the $j$-th halo has mass $M_j$ and position $\x_j$.
In particular, in our collisionless approximation
(\S\ref{ssec:pipeline_approximation}),
the electron field $\delta_e$ and galaxy field $\delta_g$ are of
the form~(\ref{eq:delta_X}), with weight functions $W_X(M,z)$ given by:
\be
W_e(M,k) = \frac{M}{\rho_m} u_M(k)
\hspace{1.5cm}
W_g(M,k) = \left\{ \begin{array}{cl}
  1/n_h & \mbox{ if } M \ge M_{\rm min} \\
  0 & \mbox{ if } M < M_{\rm min}
\end{array} \right.
\ee
where $u_M(k)$ is the Fourier-transformed density profile of a halo
of mass $M$, normalized so that $u_M(0)=1$.

\subsection{Expectation values in a fixed realization of the halo source field}

Expectation values in the halo model can be calculated in two steps.
First, we take an ``inner'' average over Poisson-sampled halos,
in a fixed realization of the source field $s(M,\x)$.
Second, we take an ``outer'' average over realizations of $s(M,\x)$,
or equivalently realizations of $\delta_{\rm lin}(\x)$ via
Eq.~(\ref{eq:s_def}).
In this section, we will analyze the inner average.
We consider an $n$-point expectation value
$\langle \delta_{X_1}(\k_1) \cdots \delta_{X_n}(\k_n) \rangle_s$,
where the suffix $\langle \cdot \rangle_s$ means that the expectation
value is taken over Poisson placements of halos, in a fixed realization
of $s(M,\x)$.

We plug in the definition~(\ref{eq:delta_X}) of $\delta_X$, to write
the expectation value as a sum over $n$-tuples of halos:
\be
\big\langle \delta_{X_1}(\k_1) \cdots \delta_{X_n}(\k_n) \big\rangle_s
= \left\langle \sum_{j_1,\cdots,j_n}
\left( \prod_{i=1}^n W_{X_i}(M_{j_i},k_i) e^{-i\k_i\cdot\x_{j_i}} \right)
\right\rangle_s
\ee
Then, as usual in the halo model, we split the sum into combinatorial terms, based on
which elements of the $n$-tuple $(j_1,\cdots,j_n)$ are equal to each other.
For example, consider the four-point function
\be
\big\langle \delta_g(\k_1) \delta_e(\k_2) \delta_g(\k_3)
\delta_e(\k_4) \big\rangle  \label{eq:diag4}
\ee
which is a subset of the six-point function~(\ref{eq:diag6}).
Writing the four-point function as a sum over halo quadruples,
we could keep terms $(j_1,j_2,j_3,j_4)$ such that
\be
j_1 = j_4 = j
\hspace{0.5cm} \mbox{and} \hspace{0.5cm}
j_2 = j_3 = j'
\hspace{0.5cm} \mbox{with $j \ne j'$}
\ee
obtaining a contribution which we will denote by $T$:
\be
T = \left\langle \sum_{j \ne j'}
  \Big( W_g(M_j,k_1) W_e(M_j,k_4) e^{-i(\k_1+\k_4)\cdot\x_j} \Big)
  \Big( W_e(M_{j'},k_2) W_g(M_{j'},k_3) e^{-i(\k_2+\k_3)\cdot\x_{j'}} \Big)
\right\rangle_s
\ee
This term $T$ is one of 7 ``two-halo'' terms which contribute to
the four-point function~(\ref{eq:diag4}), out of 15 total terms.
Physically, $T$ corresponds to summing over all quadruples
$(g_1,e_2,g_3,e_4)$ such that galaxy $g_1$ and electron $e_4$
are in one halo, and galaxy $g_2$ and electron $e_3$ are in
a different halo.
To compute $T$, we replace each sum $\sum_j$ by an integral
$\int d^3\x \, dM \, s(M,\x)$, obtaining:
\begin{align}
T &=
  \left( \int d^3\x \, dM \,
  W_g(M,k_1) W_e(M,k_4) s(M,\x) e^{-i(\k_1+\k_4)\cdot\x} \right) \nn \\
& \hspace{0.5cm} \times
  \left( \int d^3\x' \, dM' \,
  W_e(M',k_2) W_g(M',k_3) s(M',\x') e^{-i(\k_2+\k_3)\cdot\x'} \right) \nn \\
&=
  \left( \int dM \, W_g(M,k_1) W_e(M,k_4) s(M,\k_1+\k_4) \right)
  \left( \int dM' \, W_e(M',k_2) W_g(M',k_3) s(M,\k_2+\k_3) \right)
\end{align}
We now introduce diagrammatic notation, representing this equation by the diagram:
\be
T = \left(
\adjustbox{valign=c}{\begin{tikzpicture}
  \begin{feynhand}
    \vertex [ringdot] (h14) at (1.5,1) {};
    \node at (1.5,1.3) {$M$};
    \vertex (k1) at (0,1) {$\delta_g(\k_1)$};
    \vertex (k4) at (3,1) {$\delta_e(\k_4)$};
    \vertex [ringdot] (h23) at (1.5,0) {};
    \node at (1.5,0.3) {$M'$};
    \vertex (k2) at (0,0) {$\delta_e(\k_2)$};
    \vertex (k3) at (3,0) {$\delta_g(\k_3)$};
    \propag [fer] (h14) to (k1);
    \propag [fer] (h14) to (k4);
    \propag [fer] (h23) to (k2);
    \propag [fer] (h23) to (k3);
    \node at (1.5,-0.3) {$\phantom{M}$};
  \end{feynhand}
\end{tikzpicture}}
\right)  \label{eq:T_diagram}
\ee
where diagrams are translated to equations using the rules:
\be
\adjustbox{valign=c}{\begin{tikzpicture}
  \begin{feynhand}
    \vertex [ringdot] (h) at (0,0) {};
    \vertex (k1) at (-1.0,1) {};
    \vertex (k2) at (-0.6,1) {};
    \vertex (k3) at (1.0,1) {};
    \node at (0.2,0.7) {$\cdots$};
    \node at (-0.95,1.1) {$\k_1$};
    \node at (-0.55,1.1) {$\k_2$};
    \node at (1,1.1) {$\k_n$};
    \node at (0,-0.3) {$M$};
    \propag [fer] (h) to (k1);
    \propag [fer] (h) to (k2);
    \propag [fer] (h) to (k3);
  \end{feynhand}
\end{tikzpicture}}  = \int dM \, s(M,\smallsum\k_i)
\hspace{1.25cm}
\adjustbox{valign=c}{\begin{tikzpicture}
    \begin{feynhand}
      \vertex [ringdot] (a) at (0,0) {};
      \vertex (b) at (1.6,0) {};
      \node at (-0.3,0) {$M$};
      \node at (0.8,0.3) {$\delta_X(\k)$};
      \node at (0.8,-0.3) {$\phantom{\delta_X}$};
      \propag [fer] (a) to (b);
    \end{feynhand}
\end{tikzpicture}} = W_X(M,k)
\label{eq:rules_sfixed}
\ee
In general, an $n$-point correlation function
$\langle \delta_{X_i}(\k_i) \cdots \delta_{X_n}(\k_n) \rangle_s$
is the sum over all diagrams obtained using these rules.
External lines in the diagrams correspond to fields being
correlated, and vertices correspond to halos.

\subsection{Fully averaged expectation values}

The diagrammatic rules just derived in Eq.~(\ref{eq:rules_sfixed})
correspond to an expectation value $\langle \cdot \rangle_s$ over
Poisson placements of halos, in a fixed realization of the halo
source field $s(M,\x)$.
In this section, we take the ``outer'' expectation value over $s$.
We also consider $n$-point functions which contain factors of the linear
density field $\delta_{\rm lin}(\k)$, or the radial velocity $v_r(\k)$,
so that our machinery will be general enough to calculate the six-point
function~(\ref{eq:diag6}).

In general, the source function $s(M,\x)$ will depend on the halo bias model.
We will consider the simplest possibility, namely the linear bias model
$s(M,\x) = n(M) (1 + b(M) \delta_{\rm lin}(\x))$, or equivalently in
Fourier space:
\be
s(M,\k) = n(M) \Big[ (2\pi)^3 \delta^3(\k) + b(M) \delta_{\rm lin}(\k) \Big]
\label{eq:s_fourier}
\ee
Now consider a quantity which depends on the halo source field $s(M,\k)$,
such as the term $T$ from the previous section:
\be
  T
  = \int dM \, dM' \, W_g(M,k_1) W_e(M,k_4) W_e(M',k_2) W_g(M',k_3)
  \, \Big[ s(M,\k_1+\k_4) s(M,\k_2+\k_3) \Big]
\ee
To average over $s$, we replace all factors of $s$ by the RHS of
Eq.~(\ref{eq:s_fourier}), and take the expectation value over
$\delta_{\rm lin}$ using Wick's theorem. This gives:
\begin{align}
  \langle T \rangle
  &= \int dM \, dM' \, W_g(M,k_1) W_e(M,k_4) W_e(M',k_2) W_g(M',k_3) n(M) n(M') \nn \\
  & \hspace{1cm} \times \Big[ (2\pi)^6 \delta^3(\k_1+\k_2) \delta^3(\k_3+\k_4) + b(M) b(M') (2\pi)^3 \delta^3(\smallsum\k_i) \Big]
\end{align}
Diagramatically, we represent this procedure for averaging over $s$ as follows.
We start with the diagram~(\ref{eq:T_diagram}) representing $T$, in which each hollow
circle contains one factor of $s(M,\k)$.
We sum over all ways of either pairing vertices with wavy lines (representing a
Wick contraction proportional to $P_{\rm lin}$), or leaving vertices unpaired.
In the case of $T$, there are two possibilities:
\be
\adjustbox{valign=c}{\begin{tikzpicture}
  \begin{feynhand}
    \vertex [dot] (h14) at (1.5,1) {};
    \node at (1.5,1.3) {$M$};
    \vertex (k1) at (0,1) {$\delta_g(\k_1)$};
    \vertex (k4) at (3,1) {$\delta_e(\k_4)$};
    \vertex [dot] (h23) at (1.5,0) {};
    \node at (1.5,0.3) {$M'$};
    \vertex (k2) at (0,0) {$\delta_e(\k_2)$};
    \vertex (k3) at (3,0) {$\delta_g(\k_3)$};
    \propag [fer] (h14) to (k1);
    \propag [fer] (h14) to (k4);
    \propag [fer] (h23) to (k2);
    \propag [fer] (h23) to (k3);
    \node at (1.5,-0.3) {$\phantom{M}$};
  \end{feynhand}
 \end{tikzpicture}}
\hspace{1cm}
\adjustbox{valign=c}{\begin{tikzpicture}
  \begin{feynhand}
    \vertex [dot] (h14) at (1.5,1) {};
    \node at (1.5,1.3) {$M$};
    \vertex (k1) at (0,1) {$\delta_g(\k_1)$};
    \vertex (k4) at (3,1) {$\delta_e(\k_4)$};
    \vertex [dot] (h23) at (1.5,0) {};
    \node at (1.8,0.3) {$M'$};
    \vertex (k2) at (0,0) {$\delta_e(\k_2)$};
    \vertex (k3) at (3,0) {$\delta_g(\k_3)$};
    \propag [fer] (h14) to (k1);
    \propag [fer] (h14) to (k4);
    \propag [fer] (h23) to (k2);
    \propag [fer] (h23) to (k3);
    \propag [pho] (h14) to (h23);
    \node at (1.5,-0.3) {$\phantom{M}$};
  \end{feynhand}
\end{tikzpicture}}
\ee
where the diagrams are interpreted using the following diagrammatic rules:
\begin{align}
\adjustbox{valign=c}{\begin{tikzpicture}
  \begin{feynhand}
    \vertex [dot] (h) at (0,0) {};
    \vertex (k1) at (-1.0,1) {};
    \vertex (k2) at (-0.6,1) {};
    \vertex (k3) at (1.0,1) {};
    \node at (0.2,0.7) {$\cdots$};
    \node at (-0.95,1.1) {$\k_1$};
    \node at (-0.55,1.1) {$\k_2$};
    \node at (1,1.1) {$\k_n$};
    \node at (0,-0.3) {$M$};
    \propag [fer] (h) to (k1);
    \propag [fer] (h) to (k2);
    \propag [fer] (h) to (k3);
  \end{feynhand}
\end{tikzpicture}} &= \int dM \, n(M) \, (2\pi)^3 \delta^3(\smallsum\k_i)
&  
\adjustbox{valign=c}{\begin{tikzpicture}
    \begin{feynhand}
      \vertex [dot] (a) at (0,0) {};
      \vertex (b) at (1.6,0) {};
      \node at (-0.3,0) {$M$};
      \node at (0.8,0.3) {$\delta_X(\k)$};
      \node at (0.8,-0.3) {$\phantom{\delta_X}$};
      \propag [fer] (a) to (b);
    \end{feynhand}
\end{tikzpicture}} = W_X(M,k)
\nn \\  
\adjustbox{valign=c}{\begin{tikzpicture}
  \begin{feynhand}
    \vertex [dot] (h) at (0,0) {};
    \vertex (k1) at (-1.0,1) {};
    \vertex (k2) at (-0.6,1) {};
    \vertex (k3) at (1.0,1) {};
    \vertex (q) at (0,-0.8) {};
    \node at (0.2,0.7) {$\cdots$};
    \node at (-0.95,1.1) {$\k_1$};
    \node at (-0.55,1.1) {$\k_2$};
    \node at (1,1.1) {$\k_n$};
    \node at (0.3,-0.5) {$\q$};
    \node at (-0.3,0) {$M$};
    \propag [fer] (h) to (k1);
    \propag [fer] (h) to (k2);
    \propag [fer] (h) to (k3);
    \propag [pho] (h) to (q);
  \end{feynhand}
\end{tikzpicture}} &= \int dM \, n(M) b(M) \, (2\pi)^3 \delta^3(\q+\smallsum\k_i)
&  
\adjustbox{valign=c}{\begin{tikzpicture}
    \begin{feynhand}
      \vertex (a) at (0,0) {};
      \vertex (b) at (1.6,0) {};
      \node at (0.8,0.3) {$\q$};
      \node at (0.8,-0.3) {$\phantom{\q}$};
      \propag [pho] (a) to (b);
    \end{feynhand}
\end{tikzpicture}} = \int \frac{d^3\q}{(2\pi)^3} \, P_{\rm lin}(q)
\label{eq:rules_full}
\end{align}
Note that we use hollow vertices in diagrams where $s$ is not averaged
(Eq.~(\ref{eq:rules_sfixed})), and solid vertices in diagrams where $s$
is averaged (Eq.~(\ref{eq:rules_full})).

A $n$-point expectation value of the form
$\langle \delta_{X_i}(\k_1) \cdots \delta_{X_n}(\k_n) \rangle$
may be computed by enumerating all diagrams, using the preceding
diagrammatric rules.
An $n$-point function which also contains factors of $\delta_{\rm lin}(\k)$
or $v_r(\k)$, such as the six-point function~(\ref{eq:diag6}),
can be represented diagrammatically by adding the following
external lines:
\be
\adjustbox{valign=c}{\begin{tikzpicture}
    \begin{feynhand}
      \vertex (a) at (0,0) {};
      \vertex (b) at (1.6,0) {};
      \node at (0.8,0.3) {$\delta_{\rm lin}(\k)$};
      \node at (0.8,-0.3) {$\phantom{\delta_{\rm lin}(\k)}$};
      \propag [pho] (a) to (b);
    \end{feynhand}
\end{tikzpicture}} = P_{\rm lin}(k)
\hspace{2cm}
\adjustbox{valign=c}{\begin{tikzpicture}
    \begin{feynhand}
      \vertex (a) at (0,0) {};
      \vertex (b) at (1.6,0) {};
      \node at (0.8,0.3) {$v_r(\k)$};
      \node at (0.8,-0.3) {$\phantom{v_r(\k)}$};
      \propag [pho] (a) to (b);
    \end{feynhand}
\end{tikzpicture}} = \frac{ik_r}{k} P_{mv}(k)
\label{eq:rules_lin}
\ee
where we have assumed that $v_r(\k)$ is evaluated on a linear scale,
so that $v_r(\k) = (ik_r/k) v(\k) = (ik_r/k) (faH/k) \delta_{\rm lin}(\k)$.

\subsection{The six-point function $\langle \delta_g^2 \delta_e^2 v_r^2 \rangle$}

We calculate the non-Gaussian six-point function~(\ref{eq:diag6})
using the diagrammatic rules in Eqs.~(\ref{eq:rules_full}),~(\ref{eq:rules_lin}).
Up to permutations of external legs, there are five possible diagrams:
\begin{align}
  D_1 &=
  \left( \adjustbox{valign=c}{\begin{tikzpicture}
      \begin{feynhand}
        \vertex[dot] (ha) at (1,1) {};
        \vertex[dot] (hb) at (4.5,1) {};
        \vertex (k1) at (0,0) {$\delta_g(\k_1)$};
        \vertex (k2) at (2,0) {$\delta_e(\k_2)$};
        \vertex (k5) at (1,2) {$v_r(\k_5)$};
        \vertex (k3) at (3.5,0) {$\delta_g(\k_3)$};
        \vertex (k4) at (5.5,0) {$\delta_e(\k_4)$};
        \vertex (k6) at (4.5,2) {$v_r(\k_6)$};
        \propag [fer] (ha) to (k1);
        \propag [fer] (ha) to (k2);
        \propag [pho] (ha) to (k5);
        \propag [fer] (hb) to (k3);
        \propag [fer] (hb) to (k4);
        \propag [pho] (hb) to (k6);
      \end{feynhand}
  \end{tikzpicture}} \right)
& 
  D_2 &=
    \left( \adjustbox{valign=c}{\begin{tikzpicture}
    \begin{feynhand}
      \vertex[dot] (h1) at (10.5,1) {};
      \vertex[dot] (h2) at (13,1.25) {};
      \vertex (k2) at (9,0) {$\delta_e(\k_2)$};
      \vertex (k3) at (9,1) {$\delta_g(\k_3)$};
      \vertex (k4) at (9,2) {$\delta_e(\k_4)$};
      \vertex (k6) at (12,1) {};
      \vertex (k1) at (13,0) {$\delta_g(\k_1)$};
      \vertex (k5) at (13,2) {$v_r(\k_5)$};
      \node at (11.5,1.35) {$v_r(\k_6)$};
      \propag [fer] (h1) to (k2);
      \propag [fer] (h1) to (k3);
      \propag [fer] (h1) to (k4);
      \propag [pho] (h1) to (k6);
      \propag [fer] (h2) to (k1);
      \propag [pho] (h2) to (k5);
    \end{feynhand}
  \end{tikzpicture}} \right)
\nn \\ 
  D_3 &=
  \left( \adjustbox{valign=c}{\begin{tikzpicture}
    \begin{feynhand}
      \vertex[dot] (ha) at (1,1) {};
      \vertex (ka1) at (0,0) {$\delta_g(\k_1)$};
      \vertex (ka2) at (0,2) {$\delta_e(\k_2)$};
      \vertex (ka3) at (2,0) {$\delta_g(\k_3)$};
      \vertex (ka4) at (2,2) {$\delta_e(\k_4)$};
      \vertex (k5) at (3.5,0) {$v_r(\k_5)$};
      \vertex (k6) at (3.5,2) {$v_r(\k_6)$};
      \propag [fer] (ha) to (ka1);
      \propag [fer] (ha) to (ka2);
      \propag [fer] (ha) to (ka3);
      \propag [fer] (ha) to (ka4);
      \propag [pho] (k5) to (k6);
    \end{feynhand}
  \end{tikzpicture}} \right)
& 
  D_4 &=
  \left( \adjustbox{valign=c}{\begin{tikzpicture}
    \begin{feynhand}
      \vertex[dot] (hb1) at (5,1) {};
      \vertex[dot] (hb2) at (6,1) {};
      \vertex (kb1) at (4,0) {$\delta_g(\k_1)$};
      \vertex (kb2) at (4,2) {$\delta_e(\k_2)$};
      \vertex (kb3) at (7,0) {$\delta_g(\k_3)$};
      \vertex (kb4) at (7,2) {$\delta_e(\k_4)$};
      \vertex (k5) at (8.5,0) {$v_r(\k_5)$};
      \vertex (k6) at (8.5,2) {$v_r(\k_6)$};
      \propag [fer] (hb1) to (kb1);
      \propag [fer] (hb1) to (kb2);
      \propag [fer] (hb2) to (kb3);
      \propag [fer] (hb2) to (kb4);
      \propag [pho] (hb1) to (hb2);
      \propag [pho] (k5) to (k6);
    \end{feynhand}
  \end{tikzpicture}} \right)
\nn \\ 
  D_5 &=
  \left( \adjustbox{valign=c}{\begin{tikzpicture}
    \begin{feynhand}
      \vertex[dot] (hc1) at (10.5,1) {};
      \vertex[dot] (hc2) at (11.5,1) {};
      \vertex (kc1) at (9,0) {$\delta_g(\k_1)$};
      \vertex (kc2) at (9,1) {$\delta_e(\k_2)$};
      \vertex (kc3) at (9,2) {$\delta_g(\k_3)$};
      \vertex (kc4) at (12.5,1) {};
      \vertex (k5) at (13,0) {$v_r(\k_5)$};
      \vertex (k6) at (13,2) {$v_r(\k_6)$};
      \node at (12,1.35) {$\delta_e(\k_4)$};
      \propag [fer] (hc1) to (kc1);
      \propag [fer] (hc1) to (kc2);
      \propag [fer] (hc1) to (kc3);
      \propag [fer] (hc2) to (kc4);
      \propag [pho] (hc1) to (hc2);
      \propag [pho] (k5) to (k6);
    \end{feynhand}
  \end{tikzpicture}} \right)
\end{align}
In particular, there are no fully connected diagrams,
as claimed in the main text (\S\ref{ssec:n32}).
We evaluate these diagrams as follows (denoting
$\k_{i_1\cdots i_n} = (\k_{i_1} + \cdots + \k_{i_n})$):
\begin{align}
D_1 &= \left( \int dM \, n(M) b(M) W_g(M,k_1) W_e(M,k_2)
            \frac{ik_{5r}}{k_5} P_{mv}(k_5) \,
            (2\pi)^3 \delta^3(\k_{125}) \right) \nn \\
    & \hspace{1.5cm} \times \left( \int dM' \, n(M') b(M') W_g(M',k_3)
            W_e(M',k_4) \frac{ik_{6r}}{k_6} P_{mv}(k_6) \,
            (2\pi)^3 \delta^3(\k_{346}) \right) \nn \\
    &= -\beta_1(k_2) \beta_1(k_4) \frac{k_{5r}k_{6r}}{k_5k_6}
         P_{mv}(k_5) P_{mv}(k_6) \, (2\pi)^6 \delta^3(\k_{125})
         \delta^3(\k_{346}) \\
D_2 &= \left( \int dM \, n(M) b(M) W_e(M,k_2) W_g(M,k_3)
         W_e(M,k_4) \frac{ik_{6r}}{k_6} P_{mv}(k_6)
         (2\pi)^3 \delta^3(\k_{2346}) \right) \nn \\
    & \hspace{1.5cm} \times \left( \int dM' \, n(M') b(M')
         W_g(M',k_1) \frac{ik_{5r}}{k_5} P_{mv}(k_5)
         (2\pi)^3 \delta^3(\k_{15}) \right) \nn \\
    &= -b \beta_2(k_2,k_4) \frac{k_{5r}k_{6r}}{k_5k_6}
         P_{mv}(k_5) P_{mv}(k_6) \, (2\pi)^6 \delta^3(\k_{2346})
         \delta^3(\k_{15}) \\
D_3 &=
  \int dM \, n(M) W_g(M,k_1) W_e(M,k_2) W_g(M,k_3) W_e(M,k_4)
    P_{v_r}(\k_5) \, (2\pi)^6 \delta^3(\k_{1234}) \delta^3(\k_{56}) \nn \\
 &= \frac{\alpha_2(k_2,k_4)}{n_h}
    P_{v_r}(\k_5) \, (2\pi)^6 \delta^3(\k_{1234}) \delta^3(\k_{56})  \\
D_4 &=
\int dM \, dM' \, n(M) n(M') W_g(M,k_1) W_e(M,k_2) W_g(M',k_3) W_e(M',k_4) \nn \\
& \hspace{1.5cm} \times
   b(M) b(M') P_{\rm lin}(\k_1+\k_2)
   P_{v_r}(\k_5) \, (2\pi)^6 \delta^3(\k_{1234}) \delta^3(\k_{56}) \nn \\
& = \beta_1(k_2) \beta_1(k_4) P_{\rm lin}(\k_1+\k_2)
    P_{v_r}(\k_5) \, (2\pi)^6 \delta^3(\k_{1234}) \delta^3(\k_{56})  \\
D_5 &= \int dM \, dM' \, n(M) n(M') W_g(M,k_1) W_e(M,k_2) W_g(M,k_3) W_e(M',k_4) \nn \\
& \hspace{1.5cm} \times
   b(M) b(M') P_{\rm lin}(k_4)
    P_{v_r}(\k_5) \, (2\pi)^6 \delta^3(\k_{1234}) \delta^3(\k_{56})  \nn \\
& = \beta_1(k_2) \beta_1(k_4) P_{\rm lin}(k_4)
    P_{v_r}(\k_5) \, (2\pi)^6 \delta^3(\k_{1234}) \delta^3(\k_{56})
\end{align}
where for each diagram, the first line on the RHS gives the result of
applying the diagrammatic rules straightforwardly, and the second line
uses the $\alpha,\beta$ notation from
Eqs.~(\ref{eq:alpha_def})--(\ref{eq:beta_prime_def}).

Comparing with the expression for the six-point function in the main
text (Eqs.~(\ref{eq:sixpt_factorization})--(\ref{eq:S4})),
the diagram $D_1$ is the
term $(Q_{\k_1\k_2\k_5} Q_{\k_3\k_4\k_6})$ on the first line of
Eq.~(\ref{eq:sixpt_factorization}). 
One can check that the other five $QQ$-terms
in~(\ref{eq:sixpt_factorization})
are obtained by permuting external legs of $D_1$.

There is a similar story for the other diagrams.
The diagram $D_2$ is the first $PR$-term on the second
line of Eq.~(\ref{eq:sixpt_factorization}), 
and the other seven $PR$-terms are obtained
by permuting external legs of $D_2$.
The diagram $D_3$ corresponds to the first $S$-term in
Eq.~(\ref{eq:sixpt_factorization}).
The next three $S$-terms in~(\ref{eq:sixpt_factorization}) correspond to the
diagram $D_4$, and diagrams obtained from $D_4$ by permuting external legs.
Finally, the last four $S$-terms in~(\ref{eq:sixpt_factorization}) correspond to $D_5$,
and diagrams obtained from $D_5$ by permuting external legs.
Putting all 22 diagrams together gives the six-point
function shown in the main text 
(Eqs.~(\ref{eq:sixpt_factorization})--(\ref{eq:S4})).
Deriving this result was the main goal of this appendix.

\subsection{Discussion and generalizations}
\label{ssec:diag_discussion}

Diagrammatic rules make some properties of the halo model more transparent.
For example,  a {\em connected} $n$-point function 
$\langle \delta_{X_1}(\k_1) \cdots \delta_{X_n}(\k_n) \rangle_c$
consists of a one-halo term, plus $(2^{n-1}-1)$ two-halo terms containing one power of
$P_{\rm lin}$, with no terms with $\ge 3$ halos. This is easy to see from the diagrammatic
rules, but not so obvious otherwise.

We have only considered the simplest version of the halo model:
linearly biased tracers of a Gaussian field.
The diagrammatic rules can be extended to generalizations of this
model as well.
We sketch a few examples, without attempting to be exhaustive.

Our assumption of linear halo bias can be generalized,
for example by a higher-order bias model of the form
$\delta_h = b \delta_{\rm lin} + b_2 \delta_{\rm lin}^2 + \cdots$.
This can be incorporated by adding new vertices to the diagrammatic
rules in Eq.~(\ref{eq:rules_full}), such as:
\be
\adjustbox{valign=c}{\begin{tikzpicture}
  \begin{feynhand}
    \vertex [dot] (h) at (0,0) {};
    \vertex (k1) at (-1.0,1) {};
    \vertex (k2) at (-0.6,1) {};
    \vertex (k3) at (1.0,1) {};
    \vertex (q1) at (-0.5,-0.8) {};
    \vertex (q2) at (0.5,-0.8) {};
    \node at (0.2,0.7) {$\cdots$};
    \node at (-0.95,1.1) {$\k_1$};
    \node at (-0.55,1.1) {$\k_2$};
    \node at (1,1.1) {$\k_n$};
    \node at (-0.05,-0.55) {$\q_1$};
    \node at (0.6,-0.55) {$\q_2$};
    \node at (-0.5,0) {$M$};
    \propag [fer] (h) to (k1);
    \propag [fer] (h) to (k2);
    \propag [fer] (h) to (k3);
    \propag [pho] (h) to (q1);
    \propag [pho] (h) to (q2);
  \end{feynhand}
\end{tikzpicture}} = \int dM \, n(M) b_2(M) \, (2\pi)^3 \delta^3(\q_1+\q_2+\smallsum\k_i)
\ee
This would give rise to loop diagrams and renormalization, whereas
linear bias (as assumed in the main paper) only produces tree diagrams.

As another extension of the halo model, suppose that
galaxies are derived from halos using an additional
level of Poisson sampling. More precisely, assume that in a halo of
mass $M$, the number of galaxies is a Poisson random
variable with mean $N_g(M)$, and the spatial location
of each galaxy is a random variable with profile $u_g(M,k)$.
We introduce square vertices for galaxies (continuing to
denote halos by circular vertices), which are endpoints for
external legs of the form $\delta_g(\k)$.
For example, the following diagram represents a one-halo,
two-galaxy term in the three-point function
$\langle \delta_g(\k_1) \delta_g(\k_2) \delta_g(\k_3) \rangle$:
\be
\adjustbox{valign=c}{\begin{tikzpicture}
  \begin{feynhand}
  \vertex [dot] (h) at (0,0) {};
  \vertex [squaredot] (g12) at (-1.5,0) {};
  \vertex [squaredot] (g3) at (1.5,0) {};
  \vertex (k1) at (-2.5,-0.7) {};
  \vertex (k2) at (-2.5,0.7) {};
  \vertex (k3) at (2.5,0) {};
  \node at (-3,-0.7) {$\delta_g(\k_1)$};
  \node at (-3,0.7) {$\delta_g(\k_2)$};
  \node at (2,0.3) {$\delta_g(\k_3)$};
  \node at (-0.75,0.3) {$\k_1+\k_2$};
  \node at (0.75,0.3) {$\k_3$};
  \propag [fer] (h) to (g12);
  \propag [fer] (h) to (g3);
  \propag [fer] (g12) to (k1);
  \propag [fer] (g12) to (k2);
  \propag [fer] (g3) to (k3);
  \end{feynhand}
\end{tikzpicture}} =
\frac{1}{n_g^3} \int dM \, n(M) N_g(M)^2 u_g(M,\k_1+\k_2) u_g(M,\k_3)
\ee
Multiple galaxy populations (e.g.~centrals and satellites) can be
handled by introducing multiple galaxy vertex types.

Finally, the halo model is sometimes generalized by including nonlinear
evolution of the density field, rather than assuming
$s$ is proportional to $\delta_{\rm lin}$. Nonlinear evolution can
be incorporated by adding interaction vertices which couple three or more
wavy lines
($\adjustbox{valign=c}{\begin{tikzpicture} \begin{feynhand}
\vertex (a) at (0,0); \vertex (b) at (0.7,0); \propag [pho] (a) to (b);
\end{feynhand} \end{tikzpicture}}$), in a way which is familiar
from standard cosmological perturbation theory
(for a review see~\cite{Bernardeau:2001qr}).
Indeed, diagrammatic rules are frequently used for perturbative
calculations involving {\em continuous} LSS fields.
In this appendix, we have shown how to extend these rules
to include discrete fields derived by Poisson sampling,
such as halos and galaxies.
This way of enumerating combinatorial
terms in the halo model is convenient, especially for
higher-$n$ correlation functions, where the number of
terms is large.

\end{document}